\documentclass[aps,prd,a4paper,twocolumn,showpacs,showkeys,preprintnumbers,amsmath,amssymb,nofootinbib]{revtex4-2}
\usepackage{graphicx}
\usepackage{bm}
\usepackage{dcolumn}
\usepackage{color}
\usepackage{pst-plot}

\usepackage{mathrsfs} %% lagrangian L
\usepackage{txfonts}

\usepackage{footmisc} %Para citar pe de pagina
\newcommand{\comm}[1]{} %Para comentar
\usepackage{soul} %Para taxar

\renewcommand{\thesection}{\arabic{section}}

\numberwithin{equation}{section}

\let\oldsection\section% Store \section
\renewcommand{\section}{% Update \section
  \renewcommand{\theequation}{\thesection.\arabic{equation}}% Update equation number
  \oldsection}% Regular \section
\def\beq{\begin{equation}}
\def\eeq{\end{equation}}
\def\ln{\,\mbox{log}\,}

\def\Res{\,\mbox{Res}\,}
\renewcommand{\Re}{\,\mbox{Re}\,}
\renewcommand{\Im}{\,\mbox{Im}\,}

\newcommand{\eq}[1]{(\ref{#1})}
\newcommand{\n}[1]{\label{#1}}

\def\al{\alpha}
\def\be{\beta}
\def\ga{\gamma}
\def\Ga{\Gamma}

\def\de{\delta}
\def\De{\Delta}
\def\vp{\varepsilon}

\def\th{\theta}

\def\ka{\kappa}
\def\la{\lambda}

\def\ph{\varphi}
\def\om{\omega}

\def\lap{\Delta}
\def\na{\nabla}
\def\pa{\partial}

\begin{document}

\title{Higher-order regularity in  local and nonlocal quantum gravity} 

\author{Nicol\`o Burzill\`a}
\email{nburzilla@outlook.it}

\author{Breno L. Giacchini}
\email{breno@sustech.edu.cn}

\author{Tib\'{e}rio de Paula Netto}
\email{tiberio@sustech.edu.cn}

\author{Leonardo Modesto}
\email{lmodesto@sustech.edu.cn}

\affiliation{
{\small Department of Physics, Southern University of Science and Technology,}\\ 
{\small Shenzhen 518055, China}
}

\date{\small\today}

\begin{abstract} \noindent
In the present work we investigate the Newtonian limit of higher-derivative gravity theories with more than four derivatives in the action, including the non-analytic logarithmic terms resulting from one-loop quantum corrections. The first part of the paper deals with the occurrence of curvature singularities of the metric in the classical models.
It is shown that in the case of local theories, even though the curvature scalars of the metric are regular, invariants involving derivatives of curvatures can still diverge. 
Indeed, we prove that if the action contains $2n+6$ derivatives of the metric in both the scalar and the spin-2 sectors, then all the curvature-derivative invariants with at most $2n$ covariant derivatives of the curvatures are regular, while there exist scalars with $2n+2$ derivatives that are singular. The regularity of all these invariants can be achieved in some classes of nonlocal gravity theories.
In the second part of the paper, we show that the leading logarithmic quantum corrections do not change the regularity of the Newtonian limit.
Finally, we also consider the infrared limit of these solutions and verify the universality of the leading quantum correction to the potential in all the theories investigated in the paper.
\end{abstract}


\maketitle
\noindent

%%%%%%%%%%%%%%%%%%%%%%%%%%%%%%%%%%%
%%%%%%%%%%%%%%%%%%%%%%%%%%%%%%%%%%%

%%%%%%%%%%%%%%%%%%%%%%%%%%%%%%%
%%%%%%%%%%%%%%%%%%%%%%%%%%%%%%%
\section{Introduction}
\label{Sec1}

There has been an increasing interest in higher-derivative theories of gravity in recent years, especially those with more than four derivatives and weak nonlocalities. Such models are obtained by extending the Einstein--Hilbert action with curvature-squared terms such as $R \, F_0 (\Box)  \, R$ and $C_{\mu\nu\al\be} \, {F}_2(\Box) \, C^{\mu\nu\al\be}$, where $F_{0,2}(\Box)$ are analytic functions of the d'Alembertian and $C_{\mu\nu\al\be}$ denotes the Weyl tensor.
Among the motivations for this, we can mention the possibility of conciliating unitarity and renormalizability in the framework of perturbative quantum gravity. 

From one side, higher derivatives improve the behaviour of the propagator in the ultraviolet (UV) regime, which favours renormalizability. In fact, if $F_{0,2}$ are non-zero constants one has the fourth-derivative gravity, which is renormalizable~\cite{Stelle77}; while if $F_{0,2}(\Box)$ are non-trivial polynomials, the theory can be made super-renormalizable~\cite{AsoreyLopezShapiro}. We remark that, even if the metric tensor is not quantised, the renormalization of matter fields in curved space-time requires the introduction of at least four derivatives in the gravitational action~\cite{UtDW}.
It is well known, however, that local higher-derivative models usually contain ghost-like massive poles in the propagator, which violate unitarity.
In this approach, unitarity can be recovered either by projecting out the ghost-like poles of the propagator from the spectrum~\cite{ModestoShapiro16,Modesto16,AnselmiPiva1,AnselmiPiva2,AnselmiPiva3,Anselmi:2017ygm}, or avoiding them \textit{ab initio} by means of nonlocality~\cite{Krasnikov,Kuzmin,Tomboulis,Modesto12}. 

The former possibility includes the Lee--Wick gravity, which requires an action with at least six derivatives of the metric so that all the ghost-like poles of the propagator can be complex~\cite{ModestoShapiro16,Modesto16}. In the work~\cite{ModestoShapiro16} it was shown that these models are unitary at tree-level; while the generalisation of the Lee--Wick quantisation prescription~\cite{LW1,LW2,CLOP} proposed in Refs.~\cite{AnselmiPiva1,AnselmiPiva2} guarantees unitarity at any perturbative order~\cite{AnselmiPiva3}. It is worthwhile to mention that the same procedure can be applied to real ghost modes~\cite{Anselmi:2017ygm}, \textit{e.g.}, in the context of the simplest fourth-derivative gravity or to the additional ghost modes of higher-order theories.

In what concerns the nonlocal higher-derivative gravity theories, we note that if the analytic functions $F_{0,2} (\Box)$ are not polynomials and have the form
\beq
\n{nonlocal}
F_s (\Box) = \frac{ e^{H_s(\Box)} - 1 } {\Box},
\eeq
where $H_{0,2}(z)$ is an entire function, the propagator only contains the massless pole of the graviton. Therefore, these models are automatically tree-level ghost-free; they can also be (super-)renormalizable, depending on the choice of the functions $H_{0,2}(z)$~\cite{Tomboulis,Modesto12}. For a discussion on solutions and stability issues in nonlocal gravity models, see, \textit{e.g.}, Refs.~\cite{Li:2015bqa,Calcagni:2010ab,Calcagni:2018pro,Calcagni:2017sov,Briscese:2019rii,Briscese:2018bny}.

In the present work, we investigate general properties of the Newtonian limit of a generic higher-derivative gravitational theory, including the leading logarithm quantum corrections, with particular focus on the occurrence of curvature singularities and on the far infra-red (IR) behaviour. Accordingly, the gravity effective action of our interest has the general structure
\begin{equation} 
\begin{split}
\label{act}
\Ga %_{\text{vac}} 
=  & - \frac{1}{\varkappa^2} \int d^4 x \sqrt{|g|} \,
\Big\{ 2  R 
 + \, C_{\mu\nu\al\be} \, \mathscr{F}_2(\Box) \, C^{\mu\nu\al\be}
\\
&
- \tfrac13 \, R \, \mathscr{F}_0 (\Box)  \, R
\Big\},
\end{split} 
\end{equation}
where $C_{\mu\nu\al\be}$ denotes the Weyl tensor, $\varkappa^2 =  32 \pi G$ and the form factors $\mathscr{F}_s$
have the form
\begin{equation} 
\label{FormFac}
\mathscr{F}_s (\Box) =  F_s(\Box) + \beta_s \ln ( \Box/\mu_s^2 ), \qquad s = 0,2 . 
\end{equation}
Here, $\beta_s$ are constants, $\mu_s$ are renormalization group invariant scales and $F_s (\Box)$ are analytic functions of the d'Alembertian. As discussed above, the choice of $F_s (\Box)$ corresponds to the definition of the higher-derivative sector of the classical action.
In the formula~\eqref{act} we do not write terms that are irrelevant to the weak-field limit, such as the cosmological constant, superficial terms, and $O(R^3)$-structures.

The leading logarithmic quantum corrections are related to the quantities $\beta_s$, which depend on the field content of a given quantum field theory and can be calculated (for the results of standard matter fields, see, \textit{e.g.},~\cite{BirDav}). Regarding quantum gravity contributions, in the fourth-derivative gravity the beta functions of the curvature-squared terms are unambiguous~\cite{AvraBavi85,Shapiro:1994ww}, while those in general relativity are gauge- and parametrization-dependent---an issue which can be solved by using the Vilkovisky--DeWitt formalism~\cite{Vil-unicEA,DeWitt-ea}\footnote{See~\cite{Ohta:2016npm,Goncalves:2017jxq,Giacchini:2020dhv} for a discussion about the dependence on the field parametrization in quantum general relativity, and~\cite{Giacchini:2020zrl} for a recent application of the Vilkovisky--DeWitt formalism in effective quantum gravity.}. In the local super-renormalizable models the beta functions are gauge-independent and, if the degree of the polynomial $F_s (\Box)$ is at least three, they are one-loop exact~\cite{AsoreyLopezShapiro}. However, since here we aim for general results, we leave these parameters arbitrary.

It is useful to recall some results for classical theories ({\it i.e.}, when $\be_{s} \equiv 0$) in the Newtonian limit.
For the fourth-derivative gravity, the (modified) Newtonian potential is finite at $r=0$~\cite{Stelle77}, but the curvature invariants still have singularities~\cite{Stelle78}. The situation is completely different when $F_s$ is a non-trivial polynomial; in this case, the Newtonian-limit metric is not only finite~\cite{Newton-MNS,Newton-BLG}, but all the invariants built only with curvature and metric tensors are regular~\cite{BreTib1}.  Also, for a plethora of choices of the entire functions $H_s(z)$ within the nonlocal models~\eq{nonlocal} one meets the same situation of the polynomial gravity; for a detailed discussion, see Ref.~\cite{BreTib2}. One of the present work goals is to investigate if the insertion of the logarithmic terms in~\eqref{FormFac} either improves or spoils the aforementioned results.

Actually, the generalisation carried out in this paper is threefold. First, in what concerns the conditions for the regularity of the curvature invariants, here we also consider the scalars build with derivatives of the curvatures; and it is proven that these quantities can still diverge in local higher-derivative gravity models. Then, a general characterization of theories that have a regular Newtonian limit is presented, including the case of non-analytic form factors such as~\eqref{FormFac}. The quantum corrections are treated in two different ways, namely, as the full resummation of the one-loop 1-particle irreducible dressed propagator, like in~\eqref{act}, and as the first order
correction to the 2-point correlation function, which comprises a perturbative expansion on $\be_s$.
Finally, the IR limit is also discussed, and it is shown that it has a universal behaviour related to the quantum logarithmic corrections. Due to the difference between the case of fourth-derivative gravity and the other higher-derivative models, we hereby only discuss the latter one, addressing the former case in the parallel work~\cite{Nos4der}.

The paper is organized as follows. In Sec.~\ref{Sec2} we briefly review the Newtonian limit of the higher-derivative gravity model~\eqref{act}, while in Sec.~\ref{Sec3} we present a theorem on the conditions that the metric potentials should fulfil to regularise the scalars involving derivatives of the curvatures. In Sec.~\ref{Sec4} we give two explicit examples of classical theories that satisfy the assumptions of the theorem, namely, the polynomial-derivative gravity (with simple poles in the propagator) and a nonlocal gravity model. The discussion is extended in the Sec.~\ref{Sec5}, where we characterise a large family of local and nonlocal gravity models satisfying the conditions of the theorem; this analysis also includes non-analytic quantum corrections. Finally, in Sec.~\ref{Sec6} we derive results considering a perturbative expansion of the metric potentials in the quantum-correction parameter $\be_s$. Some general results are presented, especially in the IR limit; and the quantum correction to the Newtonian potential for two specific models are explicitly evaluated: the polynomial-derivative gravity with simple poles in the propagator, and the simplest nonlocal ghost-free gravity. The results are summarised in Sec.~\ref{Sec7}, where we also draw our conclusions.

%%%%%%%%%%%%%%%%%%%%%%%%%%%%%%%%%%%
%%%%%%%%%%%%%%%%%%%%%%%%%%%%%%%%%%%

\section{Newtonian limit}
\label{Sec2}

In the weak-field approximation, we consider metric fluctuations around Minkowski space-time,
\beq
\n{mli}
g_{\mu\nu} = \eta_{\mu\nu} +  h_{\mu\nu}
\,
\eeq
and expand the action~\eqref{act} up to second order on the field~$h_{\mu\nu}$. 
The quadratic part of the action \eq{act} reads
\begin{equation}
\n{bili-EH}
\Ga^{(2)} 
= -\frac{1}{2 \varkappa^2 }  \int d^4x \, h_{\mu\nu} \, H^{\mu\nu,\al\be} \, h_{\al\be}
,
\end{equation}
where
\begin{equation}
\begin{split}
\n{He}
&
H^{\mu\nu,\al\be} =  
 f_2 (\Box) \, [\de^{\mu\nu,\al\be}  \Box
- \big(\eta^{\al(\mu} \pa^{\nu)} \pa^\be + \eta^{\be(\mu} \pa^{\nu)} \pa^\al \big)] 
\\
&
-  \frac13 [f_2 (\Box) + 2 f_0 (\Box)] \,  [\eta^{\mu\nu} \eta^{\al\be}  \Box  
-  \big( \eta^{\mu\nu} \pa^\al \pa^\be + \eta^{\al\be} \pa^\mu \pa^\nu \big)]
\\
& + \frac23 [f_2 (\Box) -  f_0 (\Box)] \frac{\pa^\mu \pa^\nu \pa^\al \pa^\be}{\Box}
,
\end{split}
\end{equation}
and the functions $f_s (z)$ are defined as
\beq \label{efizinho}
f_s (\Box) = 1 + \mathscr{F}_s (\Box) \Box
.
\eeq

The interaction between gravity and matter is introduced via the matter action
\beq
S_{\text{m}} = - \frac{1}{2 } \int d^4x \, T^{\mu\nu} \, h_{\mu\nu},
\eeq
where $T^{\mu\nu}$ is the energy-momentum tensor in the flat space-time, such that the principle of least action 
\beq
\de ( \Ga^{(2)} + S_{\text{m}}) = 0 
\eeq 
yields the equations of motion,
\beq
\label{EOM}
H^{\mu\nu,\al\be} \, h_{\al\be} = - \frac{\varkappa^2}{2}  T^{\mu\nu}.
\eeq

%%%%%%%%%%%%%%%%%%%%
As our interest is in the Newtonian limit, we shall consider the metric associated with a point-like mass in rest, whose energy-momentum tensor reads 
\begin{equation}
\n{T}
T_{\mu\nu} \,=\, \rho \, \de_\mu^0 \, \de_\nu^0 \, 
\quad \text{with} \quad \rho( \vec{r} ) \, = \, M   \de^{(3)} ( \vec{r} ).
\end{equation}
In isotropic Cartesian coordinates we have the line element
\begin{equation} \label{Metric}
ds^2 =  (1+ 2 \ph) dt^2 - (1 - 2 \psi) (dx^2+dy^2+dz^2),
\end{equation}
where $\ph = \ph(r)\,$ and $\psi = \psi(r)\,$ are the Newtonian-limit
potentials and $r \,=\, \sqrt{x^2+y^2+z^2}\,$. Writing the metric potentials in the form
\beq
\label{Chi02-Def}
\ph = \frac{1}{3} (2 \chi_2 + \chi_0) \, , \qquad \psi = \frac{1}{3} (\chi_2 - \chi_0) \,  ,
\eeq
it is possible to show that the auxiliary potentials $\chi_{0,2}$ are the solutions of~\cite{BreTib1}
\beq
\label{EqPot}
f_s (-\lap) \lap \chi_s  =   \ka_s \, \rho %\,\qquad (s=0,2) ,
\, ,
\eeq
with 
\beq
\ka_s = 2 \pi G \left(  3s - 2 \right) 
.
\eeq
%%%%%%%%%%%%%%%%%%%%

One of the benefits of working with these auxiliary potentials is the explicit separation of the contributions owed to the scalar and spin-2 degrees of freedom. In fact, to the Newtonian limit, the relevant part  of the
propagator associated to~\eqref{He} is given by
\beq
\label{prop}
G_{\mu\nu\al\be} (k)
=
 \frac{P^{(2)}_{\mu\nu\al\be}}{ k^2 f_2(k^2)}
-  \frac{P^{(0-s)}_{\mu\nu\al\be} }{ 2 k^2 f_0(k^2)}
,
\eeq
where $P^{(2)}$ and $P^{(0-s)}$ are the spin-$2$ and spin-$0$ projectors~\cite{Barnes-Rivers}, $k^2 = k_\mu k^\mu$ and we used Euclidean signature. As the massive poles of the propagator are defined by the zeros of the functions $f_{0,2} (z)$, the potential $\chi_s$ only depends on the spin-$s$ sector of the theory. Also, the overall structure of the equations defining these potentials is essentially the same, see~\eqref{EqPot}, which makes it possible to derive general results based on certain particular characteristics of the functions $f_{s} (z)$, \textit{e.g.}, by means of the effective source formalism of Sec.~\ref{source-gen} (see also~\cite{BreTib2}). 
Once the expressions for both $\chi_{0,2}$ are obtained, the potentials $\ph$ and $\psi$ can be recovered as a linear combination of them, through~\eqref{Chi02-Def}. In this sense, one can work with the spin-$s$ potentials without loss of generality.

%%%%%%%%%%%%%%%%%%%%%%%%%%%%%%%%%%%
%%%%%%%%%%%%%%%%%%%%%%%%%%%%%%%%%%% 

\section{Regularity conditions in the Newtonian limit}
\label{Sec3} 

It is widely known that the regularity of a given metric does not imply in the absence of curvature singularities. For example, consider the following curvature invariants associated to~\eqref{Metric},
\beq \label{Krets}
R_{\mu\nu\al\be}^2 = 4 \, \big[  (\De \psi)^2 + (\pa_i \pa_j \ph)^2 +  ( \pa_i \pa_j \psi)^2 \big] 
,
\eeq
\beq
\begin{split}
R_{\mu\nu}^2 =& \,\, (\De \ph)^2 + 5 (\De \psi)^2 - 2 \De \ph \De \psi
+ (\pa_i \pa_j \ph)^2 
\\
&
+ (\pa_i \pa_j \psi)^2  - 2 \pa_i \pa_j \ph \pa_i \pa_j \psi  
,   
\end{split}
\eeq
\beq
\n{Rci0}
C_{\mu\nu\al\be}^2 =  2 ( \pa_i \pa_j \chi_2)^2 - \frac{2}{3} (\De \chi_2)^2
\quad \mbox{ and } \quad
R = 2 \De \chi_0.
\eeq
In view of \eq{Chi02-Def} one can say that the absence of singularities in the scalars above is related to the regularity of the quantities\footnote{Throughout this work we use the prime and superscript notation to denote differentiation with respect to $r$.}
\begin{align}
\n{S1}
\De \chi_s &= \chi_s^{\prime\prime } + \frac{ 2  \chi_s^{\prime}}{r}
,
\\
\n{S2}
\partial_i \partial_j \chi_{s_1} \partial_i \partial_j \chi_{s_2} 
& = \chi_{s_1}^{\prime\prime} \chi_{s_2}^{\prime\prime} + \frac{2}{r^2} \, \chi_{s_1}^{\prime} \chi_{s_2}^{\prime}
,
\end{align}
where we changed to spherical coordinates in the spatial sector.
Therefore, the existence of the limits
\beq \label{RegCond}
\lim_{r \rightarrow 0} \, \chi_s^{\prime\prime}(r) 
\qquad \mbox{and} \qquad
\lim_{r \rightarrow 0} \, \frac{\chi_s^\prime(r)}{r} 
\eeq
is a sufficient condition for avoiding curvature singularities at $r=0$ in the invariants~\eqref{Krets}--\eqref{Rci0}. These conditions depend on the derivatives of the potentials $\chi_s$, not only on their finiteness; this is why there are curvature singularities in the fourth-derivative gravity although the potentials are bounded~\cite{Stelle77,Stelle78}.

It turns out that the existence of the limits~\eqref{RegCond} also ensures the regularity of the higher-order scalars of the type $\mathcal{R}^n$, formed by the contraction of an arbitrary number $n$ of curvature tensors. In fact, by dimensional arguments, such object depends solely on combinations of products of $\chi_s^{\prime\prime}$ and $\chi_s^\prime/ r$. 

Nevertheless, if one aims to regularise not only the invariants of the type $\mathcal{R}^n$, but also those involving derivatives of the curvatures, the potentials should fulfil additional conditions. This can be readily seen from the evaluation of $\Box^n R$ (for an arbitrary $n$), which, according to \eq{Rci0}, is given by 
\beq
\label{LapN_R}
\Box^n R =  2 \lap^{n+1} \chi_0.
\eeq 
For a generic central function $\pi(r)$ one has
\beq 
\label{LapN}
\lap^{n+1} \pi(r)  = \pi ^{(2n+2)}(r) + \frac{2(n+1) }{r} \pi ^{(2n+1)} (r)
.
\eeq
Therefore, to regularise the scalar  $\Box^n R$ it suffices to have a potential $\chi_0(r)$ of class $C^{2n+2}$ such that there exists the limit
\beq \label{NNumber}
\lim_{r \to 0}  \frac{\chi_0^{(2n+1)}(r)}{r}   .
\eeq
(Notice that the condition~\eqref{RegCond} is the particular case $n=0$.)
In addition, if $\chi_0(r)$ is of class $C^{2n+2}$, but~\eqref{NNumber} diverges, then $\Box^n R$ is not regular.

It is straightforward to verify that if the same conditions hold also for $\chi_2$, then the invariants of type $\Box^n \mathcal{R}^k$, for any integer $k\geqslant 1$, are finite at $r=0$. Indeed, the scalars $\mathcal{R}^k$ are built only with combinations of products of $\chi_s^{\prime\prime}$ and $\chi_s^\prime \, r^{-1}$, thus $\Box^n \mathcal{R}^k$ is formed by sums of terms $\De^n [ (\chi_{s_1}^{\prime\prime})^i (\chi_{s_2}^\prime \, r^{-1})^j ] $ for some powers $i,j$ such that $i+j=k$. Provided that the derivatives of odd order of both potentials $\chi_{0}$ and $\chi_{2}$ vanish at least up to (including) the $(2n+1)$-th order, the Taylor representation of 
$(\chi_{s_1}^{\prime\prime})^i (\chi_{s_2}^\prime \, r^{-1})^j$ has no term with odd power $r^{2\ell-1}$ for $\ell \leqslant n$. Then,  $[(\chi_{s_1}^{\prime\prime})^i (\chi_{s_2}^\prime \, r^{-1})^j]^{(2n-1)}\sim r$ and it follows from~\eqref{LapN} that
$\Box^n \mathcal{R}^k$ is regular too, for any $k \geqslant 1$. 

To establish more general results, let us define the set $\mathscr{I}_{2n}$ of all the scalars constructed with curvature tensors and their derivatives, with the restriction that the maximum number of derivatives of curvatures is $2n$. For example, 
$\mathscr{I}_{0} = \lbrace \mathcal{R}^k ; k \geqslant 1 \rbrace$,
while $\Box \mathcal{R}^k$ and $(\nabla_{\mu}{R}_{\al\be})^2$ belong to $\mathscr{I}_{2}$. Accordingly, it is clear that $\mathscr{I}_{2n} \supset \mathscr{I}_{2(n-1)} \supset \cdots \supset \mathscr{I}_{0}$. It is also useful to introduce the definition of \textit{order of regularity} of a function, as follows.

\textbf{Definition.} Given a function $\pi:[0,\infty)\to\mathbb{R}$ and an integer $p\geqslant 0$, we shall say that $p$ is the \textit{order of regularity} of $\pi$ 
if:
\begin{itemize}
\item[(i)] $\pi(r)$ is at least $2p$-times differentiable on $[0,\infty)$ and $\pi^{(2p)}(r)$ is continuous.
\item[(ii)] If $p\geqslant 1$, the  first $p$  odd-order derivatives of $\pi(r)$ vanish as $r \to 0$, namely
$$
0 \leqslant n \leqslant p-1 \quad \Longrightarrow \quad \lim_{r \to 0} \, \pi^{(2n+1)}(r) = 0 .
$$
\end{itemize}
If these conditions hold we shall also say that the function $\pi(r)$ is \textit{$p$-regular}. 

In terms of this definition, a continuous function which is regular at $\, r=0 \,$ is 0-regular, while the limits in Eq.~\eqref{RegCond} characterise the 1-regularity of a function at least twice continuously differentiable\footnote{The definition of ``finiteness'' (the standard notion of regularity) is equivalent to ``$0$-regularity''. Therefore, throughout this work, we shall simply say ``regularity'' instead of ``0-regularity'' without ambiguity in interpretation. This is in contrast to the definition of ``regularity'' adopted in~\cite{Frolov:Poly,BreTib1,BreTib2}, which coincides to what here we call ``$1$-regularity''.}.
Having Taylor's theorem in mind, one can say that a real function $\pi(r)$ is $p$-regular if the first $p$ odd-order coefficients of its Taylor polynomial around $r=0$ are zero. In this sense, an analytic function $\pi(r)$ is $\infty$-regular if and only if it is an even function. Moreover, the condition~(ii) of the definition is equivalent to say that $\pi^{(2n+1)}(r) \underset{r \to 0}{\longrightarrow} 0$ at least linearly. 

According to the discussion presented here, if the potentials $\chi_{0,2}$ are $(n+1)$-regular then there exist regular scalars with $2n$ derivatives of the curvatures. A stronger result is stated as the following theorem, whose proof we postpone to the Appendix.

\textbf{Theorem.} Given an integer $n\geqslant 0$, a sufficient condition for the regularity of all the elements in $\mathscr{I}_{2n}$ is that the potentials $\chi_0$ and $\chi_2$ are $(n+1)$-regular.

Most of the discussions in the literature on higher-derivative gravity has been focused on the regularisation of the invariants in $\mathscr{I}_{0}$ (see, \textit{e.g.},~\cite{BreTib1,BreTib2,Frolov:Poly} and references therein). One of the goals in the present work is to extend the characterisation of regular models in the Newtonian limit beyond the simplest $0$-regularity. In this spirit, in the next section we characterise the local classical higher-derivative gravity models for which the set $\mathscr{I}_{2n}$ (for a given $n>0$) only contains non-singular scalars; while in the following sections we extend considerations to the cases involving leading logarithmic quantum corrections as well as classical nonlocal gravity models.

%%%%%%%%%%%%%%%%%%%%%%%%%%%%%%%%%%%
%%%%%%%%%%%%%%%%%%%%%%%%%%%%%%%%%%%

\section{Higher-order regularity in classical polynomial-derivative gravity models}
\label{class-ex} 
\label{Sec4}

{\it 
Summary of the section: we show that if a local gravitational model has $2(N+1)$ derivatives in the spin-$s$ sector, then the potential $\chi_s(r)$ is $(N-1)$-regular, but it is not $N$-regular.
According to last section's theorem, it means that all invariants containing up to $2(N-2)$ covariant derivatives of the curvature tensors are singularity-free at $r = 0$.
As an anticipation of Sec.~\ref{Sec5}, we also give an example of a nonlocal theory for which the potentials are $\infty$-regular.
}\\

Given a function $f(z)$, the solution of~\eqref{EqPot} can be reduced to a quadrature by means of the three-dimensional Fourier or the Laplace transform methods (see, \textit{e.g.},~\cite{Newton-MNS,Newton-BLG,BreTib1,Frolov:Exp,Frolov:Poly,Buoninfante:2020qud}). In the first case, it is possible to integrate over the angular coordinates of the three-vector $\, \vec{ k} \,$, the result is:
\beq \label{Chi-Int}
\chi (r) = - \frac{\ka M}{ 2\pi^2 r} \int_0^\infty \frac{dk}{k} 
\frac{\sin (kr)}{f({ k}^2)} 
, \qquad k = | \vec{ k} |.
\eeq
Notice that we dropped the $s$-label for the sake of simplicity.

In this section we assume that $f(z)$ is a real polynomial of degree\footnote{The case $N=1$ is not considered here for it corresponds to the strictly renormalizable fourth-derivative gravity~\cite{Stelle77}, which contains singularities already in~$\mathscr{I}_{0}$.} $N>1$, which corresponds to the local super-renormalizable models of Ref.~\cite{AsoreyLopezShapiro}.
Comparing Eqs.~\eq{act} and \eq{efizinho} we see that, for this choice, the gravitational action contains $2(N+1)$ derivatives of the metric tensor. For simplicity, here we restrict considerations to the case in which the equation $f(z)=0$ has $N$ simple roots $z=-m_i^2$, with $i=1,...,N$. We allow, however, the occurrence of complex roots,  which can only appear in conjugate pairings owed to the fundamental theorem of algebra\footnote{Remember that here we consider that $f(z)$ is a real polynomial so that the action is also real and polynomial in derivatives.}. Moreover, in order to avoid tachyons in the spectrum it is assumed that $\text{Re} (m_i^2) > 0$.
The scenario with complex roots is of greatest interest from the physical viewpoint. Indeed, such models correspond to the class of Lee--Wick gravity~\cite{ModestoShapiro16,Modesto16} in which the conflict between unitarity and renormalizability is solved if the ghost degrees of freedom are quantized {\it \`a la} Lee--Wick with the prescription given in~\cite{AnselmiPiva1,AnselmiPiva2}, but without introducing any extra fictitious scale.

Under these conditions, and recalling that $f(0)=1$, the polynomial $f(z)$ can be factored as
\beq
\n{fz-poly}
f(z) = \prod_{i=1}^N \frac{z + m_i^2}{m_i^2} .
\eeq
To solve the integral \eq{Chi-Int} we can start applying the partial fraction decomposition,
\beq \label{PartFrac1}
\frac{1}{f(z)} = \prod_{i=1}^N \frac{m_i^2}{z + m_i^2}  = \sum_{i=1}^N C_i \frac{ m_i^2}{z + m_i^2}  ,
\eeq
where  
\beq
\n{coi}
C_i =  \prod_{\substack{j=1\\j \neq i}}^N \frac{m_j^2}{m_j^2 - m_i^2}  .
\eeq
Thus,
\beq \label{Chi-Int-pa}
\chi (r) = - \frac{\ka M}{ 2\pi^2 r}  \sum_{i=1}^N C_i \, m_i^2 \int_0^\infty \frac{dk}{k} 
\frac{  \sin (kr)}{k^2 + m_i^2} 
. 
\eeq
Performing an analytic continuation to the complex plane via $k \mapsto z \in \mathbb{C}$, it is possible to define a closed contour $C$ on the upper half-plane $\Pi^+=\lbrace z\in \mathbb{C}: \Im(z)\geqslant 0\rbrace$ with an indentation around the origin, such that the values of the integrals in \eq{Chi-Int-pa} over the real line are related to the poles inside $C$ by means of the Cauchy's residue theorem.  The result is~\cite{Newton-MNS,Newton-BLG}
\begin{equation}
\n{poly-class}
\chi (r) = - \frac{\ka M}{ 4 \pi r} \, \Bigg(  1
- \sum_{i=1}^N C_i \, e^{-m_i r}
\, \Bigg).
\end{equation}

Some general comments about this solution are in order. Even though the coefficients $C_i$ might be complex, the masses appear only in complex conjugate pairs, and the combination in \eq{poly-class} guarantees that the potential is a real-valued function~\cite{Newton-BLG}. From \eq{poly-class} it is possible to show that the presence of complex poles in the propagator yields oscillatory contributions to the Newtonian potentials, which are damped by Yukawa factors~\cite{Accioly:2016qeb}. In Ref.~\cite{BreTib1} the solution \eq{poly-class} was generalized to address the case of degenerate poles of arbitrary order---the outcome is that the solution gets new additional terms in the form of products of modified Bessel and power functions. 
Furthermore, it was proved that the potential \eq{poly-class} is 1-regular if $N>1$ (that is, in theories with at least sixth-derivatives)~\cite{BreTib1}. 

In what follows, we refine this result, showing that the potential is, actually, $(N-1)$-regular and that it cannot be regular to an order higher than this.
As mentioned above, in the explicit proof in this section we only deal with the case of simple poles in the propagator; the most general case is postponed to the next section.

To show that all the first $N$ odd-order Taylor coefficients of the potential \eq{poly-class} are null, let us start by writing the series explicitly, namely
\begin{equation} \label{Taylor-Pot}
\begin{split}
\chi (r) =  - \frac{\ka M}{ 4 \pi} \, \Bigg[  \Bigg( 1 - \sum_{i=1}^N C_i \Bigg)  \, \frac{1}{r} 
+ \sum_{k=0}^\infty \frac{(-1)^{k+1}}{(k+1)!} \, \Bigg( \sum_{i=1}^N C_i \, m_i^{k+1} \Bigg) \, r^k \Bigg] .
\end{split}
\end{equation}
Then, we use the relations that come from the partial fraction decomposition~\eqref{PartFrac1} to show that the  divergent and the aforementioned odd-power terms vanish.

Writing the \textit{r.h.s.} of~\eqref{PartFrac1} as a single fraction we get
\beq \label{PartFrac2}
\sum_{i=1}^N C_i \frac{ m_i^2}{z + m_i^2} = \frac{1}{\prod_i ( z + m_i^2)} \sum_{\ell =0}^{N-1} \al_{\ell} \, z^\ell
\eeq
where
\beq \label{AlfaB}
\al_{N-\ell} \equiv \sum_{i=1}^N C_i m_i^2 B_{\ell-1,i} 
\eeq
and $B_{\ell,i}$ is defined as the sum of all the combinations of products of distinct quantities $m_j^2$ with $j \neq i$, taken $\ell$ by $\ell$. For example,
\begin{eqnarray*}
B_{0,i}  \equiv  1 ,
&  &
B_{1,i}  =  \sum_{\substack{j=1\\j \neq i}}^N m_j^2 \, ,
\\
B_{2,i} \,\, =  \sum_{\substack{j,k=1\\j,k\neq i,\,j\neq k}}^N m_j^2 m_k^2 \, ,
& \quad \cdots  , \quad &
B_{N-1,i}  =  \prod_{\substack{j=1\\j\neq i}}^N m_j^2 .
\end{eqnarray*}
It is also useful to define the related quantities 
\beq
B_\ell \,\,\, = \, \sum_{\substack{k_1,\cdots,k_\ell=1 \\ k_i\neq k_j \forall i,j}}^N \,\, \prod_{i=1}^\ell m_{k_i}^2 , \qquad \ell=1,\cdots,N,
\eeq
which is the sum of all the combinations of products of distinct quantities $m_i^2$, taken $\ell$ by $\ell$; and $B_0 \equiv 1$. We have immediately the recursive formula
\beq
B_{\ell,i} = B_\ell - m_i^2 \, B_{\ell-1,i},
\eeq
that can be applied $\ell$ times to express $B_{\ell,i}$ in terms of the quantities $B_\ell$ solely, namely
\beq
B_{\ell,i} = \sum_{j=0}^\ell (-1)^j \, m_i^{2j} \, B_{\ell-j}.
\eeq
Therefore, Eq.~\eqref{AlfaB} can be rewritten as
\beq \label{alfa-final}
\al_{N-\ell-1} = \sum_{j=0}^{\ell} (-1)^j \, B_{\ell-j} \sum_{i=1}^N C_i m_i^{2(j+1)} ,
\eeq
for $\ell=0,\cdots,N-2$, while $\ell=N-1$ implies
\beq 
\alpha_0 = \left( \prod_{i=1}^N m_i^2 \right)  \sum_{j=1}^N C_j. 
\eeq

On the other hand, the comparison of Eqs.~\eqref{PartFrac1} and~\eqref{PartFrac2} yields
\beq \label{Mata-1}
\alpha_0 = \prod_{i=1}^N m_i^2 \quad  \Longrightarrow \quad \sum_{i=1}^N C_i = 1 \, .
\eeq
The relation~\eqref{Mata-1} is responsible for the finiteness of the potential at $r=0$, see~\eqref{Taylor-Pot}~\cite{Newton-MNS,Newton-BLG,BreTib1}. In the general polynomial model with simple poles considered here, it has the explicit form
\beq \label{416}
\sum_{i} \prod_{j \neq i}  \frac{m_j^2}{m_j^2 - m_i^2}  = 1.
\eeq
Although~\eqref{416}  can be proven to hold for any set of distinct numbers $\lbrace m_i^2 \rbrace$~\cite{Newton-BLG}, it is not necessary to work with the expression of the coefficients $C_j$ in terms of $m_i$ to verify the cancellation of the singularity, since~\eq{Mata-1} is merely a consequence of the partial fraction decomposition, as showed above.

Besides, we have $N-1$ relations of the type 
\beq \label{alphaEll}
\alpha_{N-\ell-1} = 0, \qquad \ell = 0,\cdots N-2.
\eeq
Hence, the system~\eqref{alfa-final} can be easily solved for the quantities $\sum_i C_i m_i^{2n}$ with $n=1,\cdots,N$, iteratively starting from $\ell=0$ and successively applying~\eqref{alphaEll} up to $\ell=N-1$, when we use~\eqref{Mata-1}. The result is: 
\beq \label{array1}
\sum_{i=1}^N C_i \, m_i^{2n}   =  \left\{ 
\begin{array}{l l}
0  ,  &  \text{if } \,  n =1,\cdots, N-1,\\
 (-1)^{N-1} \prod_{i=1}^N  m_i^2    , &  \text{if } \, n = N.
\end{array} \right .
\eeq
Comparison with~\eqref{Taylor-Pot} then revels that the potential is $(N-1)$-regular, but not $N$-regular.

It was already known that all the polynomial models with at least six derivatives have 1-regular potentials~\cite{BreTib1}. Here we showed that higher-order regularity can be achieved as one increases the number of derivatives in the action.
In nonlocal ghost-free models, depending on the choice of the entire function $H(z)$ the potentials can be $\infty$-regular. This is the case of the family of ghost-free gravities defined by the function
\beq
\n{GFNff}
f(k^2)=e^{(k^2/m^2)^N},
\eeq 
where $m$ is a mass parameter and $N\in \mathbb{N}$. In fact, in these theories the potential is an analytic even function \cite{Head-On,Edholm:2016hbt}, \textit{i.e.}, it can be expressed as a power series in $r^2$, ensuring the regularity of \textit{all} the curvature and curvature-derivative local invariants. For example, for $N=1$ the explicit solution for \eq{Chi-Int} is \cite{Tseytlin:1995uq} (see also~\cite{Modesto12,SiegelEtAl})
\begin{equation}
\n{Tseytlin}
\chi (r)  = - \frac{\kappa M}{4 \pi r} \, \text{erf}\left(\frac{m r}{2}\right) = 
- \frac{\kappa M m}{4 \pi^{3/2} } e^{-\frac{m^2r^2}{4}} M \left( 1, \tfrac32, \tfrac{m^2 r^2}{4} \right) ,
\end{equation}
where $M(a,b,z) = {}_1 F_1(a;b;z)$ is the Kummer's confluent hypergeometric function. Further discussion on the regularity in more general nonlocal models is carried out in the next section.

%%%%%%%%%%%%%%%%%%%%%%%%%%%%%%%%%%%
%%%%%%%%%%%%%%%%%%%%%%%%%%%%%%%%%%%

\section{Effective sources and regularity}
\label{source-gen}
\label{Sec5}

{\it 
Summary of the section: we relate the order of regularity of the potentials $\chi(r)$ to the behaviour of the propagator in the UV. This is done through the effective source formalism, and allows a characterization of higher-derivative models with a regular Newtonian limit. The consideration is very general and applies to local and nonlocal models, including some non-analytic form factors, such as the logarithmic quantum corrections.
}\\

 The inclusion of the logarithmic quantum correction in the functions $f_s$ makes the task of evaluating the Newtonian potentials~\eqref{Chi-Int} much more involved than in the case of analytic form factors. From one side, the pole structure of the integrand becomes considerably richer than its purely classical counterpart. To our best knowledge, only for the fourth-derivative gravity the detailed analysis of the structure of these poles, taking into account the one-loop logarithmic corrections, has been carried out~\cite{Calmet:2017omb}. Moreover, still in this simplest case it seems that knowing the pole's position is not very helpful in solving the corresponding integral~\eqref{Chi-Int} using Cauchy's theorem, and other methods should be applied~\cite{Nos4der}.

Even if the classical theory is ghost-free at tree-level, one may think that quantum corrections can introduce new ghost degrees of freedom in the propagator (see,~\textit{e.g.},~\cite{CountGhost}). We remark, however, that the logarithmic or other quantum corrections are only perturbative, and cannot affect the spectrum of the theory in their validity regime. This is strictly related to the perturbative unitarity based on the Cutkosky cutting rules, which are also based on the Landau's singularities of the loop amplitudes. Indeed, the singularities in the amplitudes in weakly non-local theories, defined by the form factors~\eqref{nonlocal}, are exactly the same of the local ones, as proved in~\cite{BrisceseModesto,Efimov,PiuSen,Chin:2018puw}. This is also evident when looking at the zeros of the denominator of the propagator, namely, considering the following equation,
$$
z [  1 +  \beta  e^{-H(z)}z \ln (z) ] = 0 \, ,
$$
which is the sum of all one-loop one-particle irreducible amplitudes. 
The zeros appear for $\vert \beta  e^{-H(z)} z \ln (z)\vert  \sim 1$, which is in contradiction with the one-loop approximation 
$\vert \beta  e^{-H(z)} z \ln(z) \vert < 1$. The case of local higher-derivative gravity models can be analysed in a similar way; all in all, this discussion is in agreement with the two explicit analytic solutions that we will present in Sec.~\ref{Sec6}, in which only the poles of  the classical propagator contribute to the Newtonian potential (see, for instance, Eqs.~\eqref{loop-pot} and~\eqref{Chi1Exp}). Ultimately, the presence (and the stability) of a ghost originated from quantum corrections can only be decided with non-perturbative results---see~\cite{Salam:1978fd,Tomboulis:1977jk,Tomboulis:1983sw,Antoniadis:1986tu,Johnston:1987ue} for a discussion related to this issue in the context of the fourth-derivative gravity. Let us mention that the logarithmic quantum corrections produce also an IR singularity in the propagator; this well-known fact is responsible precisely to the asymptotic $1/r^3$ behaviour of the one-loop correction to the Newtonian potential~\cite{Duff:1974ud,Donoghue:1993&94,Muzinich:1995uj,Hamber:1995cq,Dalvit:1997yc,Akhundov:1996jd,Khriplovich:2002bt,BjerrumBohr:2002kt}, which we will discuss in more details in Sec.~\ref{Sec6} (similar reasoning applies in quantum electrodynamics see, {\it e.g.},~\cite{HelayelNeto:1999ut}).

Owed to the mentioned difficulties to obtain explicit solutions for the metric potentials with 
nonlocal form factors (including those with
logarithmic corrections), our strategy to obtain general results concerning the occurrence of singularities in the Newtonian limit is to follow the approach used in Ref.~\cite{BreTib2} in terms of effective sources\footnote{See, \textit{e.g.},~\cite{Tseytlin:1995uq,Nicolini:2005vd,Nicolini:2008aj,Modesto:2010uh,Modesto12,Zhang14,Modesto-LWBH,JBoos,Buoninfante:2018rlq} for applications of the formalism to specific models.}. 
While in~\cite{BreTib2} the relation between the regularity of the source and the 1-regularity of the potential was derived under the assumption that the function $f(z)$ is analytic, here we shall extend the considerations to non-analytic form factors. Moreover, as in~\cite{BreTib2} the main concern was related to the regularity of the curvature invariants of the type $\mathcal{R}^n$, it was sufficient to prove the finiteness of the effective source. Here, in order to verify whether the conditions obtained in Sec.~\ref{Sec3} for the regularity of invariants constructed with derivatives of curvatures are also satisfied, we should take into account the source's differentiability.

The basic idea of the method is to rewrite Eq.~\eqref{EqPot} as a standard Poisson equation,
\begin{equation}
\label{EqPotSource}
\lap \chi  =   \ka \, \rho_\text{eff} \, ,
\, 
\end{equation}
with the modified source
\begin{equation}
\label{rho_def}
\rho_\text{eff}(r) = \frac{M}{2 \pi^2} \int_{0}^{\infty} dk  \frac{k \sin(kr)}{r f(k^2)} .
\end{equation}
In this way, the effect of a non-constant, continuous function $f(z)$ on the Newtonian potential can be treated as the smearing of the original $\de$-source~\eqref{T}---and the regularity properties of the potential $\chi$ can be deduced from those of the effective source $\rho_\text{eff}$. 

Indeed, if the source in~\eqref{EqPotSource} is bounded and integrable, then $\chi(r)$ is continuously differentiable. If, in addition, $\rho_\text{eff}(r)$ is locally Lipschitz continuous, the potential $\chi(r)$ is twice continuously differentiable (see, \textit{e.g.},~\cite{Book-PDE}). In particular, the existence of $\rho_\text{eff}(0)$ and $\chi^{\prime\prime}(0)$ implies that
$$
\lim_{r \rightarrow 0} \frac{\chi^\prime(r)}{r} < \infty,
$$
as it can be directly verified by writing the Laplacian in spherical coordinates,
\begin{equation} \label{Lapla}
\chi^{\prime\prime}(r) + \frac{2}{r} \, \chi^\prime(r)  =  \ka \, \rho_\text{eff}(r) \, ,
\end{equation}
and applying the limit $r\to 0$ in both sides of this equation. Hence, under these circumstances, the finiteness of the source means that the Newtonian potential $\chi(r)$ satisfies the conditions~\eqref{RegCond}, and all the curvature invariants of type $\mathcal{R}^n$  are regular. In what follows, we investigate the conditions for the finiteness and higher-order regularity of the effective source.

Assuming that the propagator does not have tachyonic poles,  $f(z)$ does not change sign for $z \in (0,\infty)$. Moreover, putting $f(0) = 1$ (see Eq.~\eqref{efizinho}), the function
\begin{equation} \label{g_def}
g(r,k) = \frac{k \sin(kr)}{r f(k^2)},
\end{equation}
in the integrand of~\eqref{rho_def}, is bounded on any compact. The integrability of~\eqref{g_def}, thus, depends on its behaviour as $k~\to~\infty$. It holds, however, that if there exists $k_0>0$ such that $f(k^2)$ grows at least as fast as $k^{4}$ for $k>k_0$, then
\begin{equation}
k> k_0 \quad \Longrightarrow \quad \vert g(r,k) \vert  \leqslant \frac{c}{k^2}
\end{equation}
for some constant $c$.
This means that $g(r,k)$ is integrable, even for $r=0$, provided $f(z)$ grows as $z^{2}$ or faster\footnote{\label{PDP_acucar} It is actually possible to refine this condition to faster than $z^{3/2 + \varepsilon}$ for $\varepsilon > 0$.} 
for sufficiently large arguments, as it can be proved using the Weierstrass test.
Under these circumstances, $\rho_\text{eff}(r)$ is integrable and finite for $r \geqslant 0$, showing that the $\de$-singularity of the original source is regularised by the higher derivatives~\cite{BreTib2}.

%%%%%%%%%%%%%%%%%%%%%%%%%%%%%%%%%%%
%%%%%%%%%%%%%%%%%%%%%%%%%%%%%%%%%%%

\subsection{1-regularity of the potential}
\label{Sec5a}

So far, we have established conditions for the finiteness of $\rho_\text{eff}$. It is also possible to prove that, if those conditions hold, $r=0$ is the global maximum of the effective source~\cite{BreTib2}, which is an intuitive idea. The shape of this maximum can, in principle, depend on the gravity model. Now we show that if $f(k^2)$ grows faster than $k^{4+\varepsilon}$ (for an arbitrary $\varepsilon>0$ and sufficiently large $k$) then
this maximum is ``smooth'' in the sense that the effective source is at least 1-regular,
\begin{equation} \label{DerRho}
\lim_{r\rightarrow 0} \rho_\text{eff}^\prime(r) = 0.
\end{equation}
In general, one cannot expect the effective source, viewed as a function on $\mathbb{R}^3$, to be of class $C^\infty$ because the original source is a $\de$-function. However, it is useful to recall that if the source is differentiable and $\rho_\text{eff}^\prime(r)$ is bounded, as it is here, then $\rho_\text{eff}$ is locally Lipschitz---and the Newtonian potential is 1-regular.

To prove Eq.~\eqref{DerRho}, notice that (see~\eqref{QuotaDerN} below)
\begin{equation} \label{QuotaDer}
\bigg\vert \frac{\pa}{\pa r}g(r,k) \bigg\vert = \frac{k}{r f(k^2)} \bigg\vert k \cos(kr) - \frac{\sin(kr)}{r} \bigg\vert \leqslant \frac{k^3}{2f(k^2)}.
\end{equation}
Therefore, if for $k$ large $f(k^2)$ grows at least as fast as $k^{4+\varepsilon}$ (for an $\varepsilon>0$), the  integral
\begin{equation} \label{IntUniConv}
\int_{0}^{\infty} dk \, \frac{\pa}{\pa r}g(r,k)
\end{equation}
converges uniformly for $r \geqslant 0$. Under these conditions $\rho_\text{eff}$ is differentiable and we can apply differentiation under the integral sign in~\eqref{rho_def}. Furthermore, since in the limit $r \to 0$ the function $\tfrac{\pa}{\pa r}g(r,k)$ converges uniformly to $0$ on any compact, the limit on $r$ can be interchanged with the integral in~\eqref{IntUniConv}. This gives~\eqref{DerRho}---provided that $f(k^2)$ asymptotically grows faster than $k^{4+\varepsilon}$. In particular, because $\rho_\text{eff}^\prime (r)$ is bounded, the effective source is Lipschitz continuous and the potential $\chi$ is 1-regular.

It remains to deal with the limiting situation of $\varepsilon = 0$, in which $f(k^2) \sim k^4$ asymptotically, like in  the sixth-derivative gravity. In this case, it is possible to define an integrable function which serves as an upper bound like in the \textit{r.h.s.} of~\eqref{QuotaDer}, but it depends on $r$. In fact, for $k$ sufficiently large, it holds
\begin{equation} \label{QuotaDer4th}
\bigg\vert \frac{\pa}{\pa r}g(r,k) \bigg\vert \leqslant \frac{2k^2}{r f(k^2)} \sim \frac{c}{r k^2} ,
\end{equation}
for some constant $c$.
Thence, the integral~\eqref{IntUniConv} converges uniformly on intervals which do not contain $r = 0$ as a limiting point, but $\lim_{r\to 0}\rho_\text{eff}^\prime(r)$ cannot be evaluated by interchanging the limit $r\to 0$ and the integral. Owed to this, these sources may have a spike in $r=0$ but, if they are still locally Lipschitz, $\chi$ is 1-regular.

For instance, the effective source for the classical sixth-derivative gravity (see Sec.~\ref{class-ex}) with a pair of simple poles with masses $m_1$ and $m_2$ can be read off from the general result of Ref.~\cite{BreTib2},
\beq
\rho_\text{eff}(r) = \frac{M m_1^2 m_2^2}{4\pi (m_2^2 - m_1^2)} \left( \frac{e^{-m_1 r} - e^{-m_2 r}}{r} \right) .
\eeq
It is straightforward to verify that $\rho_\text{eff}^\prime (r)$ is bounded (which is enough to guarantee the 1-regularity of the potential) even though it does not vanish at $r=0$,
\beq
\rho_\text{eff}^\prime(0) = - \frac{M m_1^2 m_2^2}{8\pi}.
\eeq

To prove that this qualitative result remains unchanged in the more general case with logarithmic quantum corrections and/or classical nonlocalities, one can change variables $kr \mapsto u$ ($r \neq 0$) in~\eqref{IntUniConv}, which becomes
\begin{equation}
\int_{0}^{\infty} dk \, \frac{\pa}{\pa r}g(r,k) = \int_{0}^{\infty} du \, \frac{u  \left( u \cos u - \sin u \right) }{r^4 f(u^2/r^2)} .
\end{equation}
Having assumed that $f(k^2) \sim k^4$ asymptotically, there exists a small enough $r_0$ such that $0< r < r_0$ yields
\begin{equation} \label{ultima}
\begin{split}
\bigg\vert \int_{0}^{\infty} dk \, \frac{\pa}{\pa r}g(r,k) \bigg\vert \leqslant & \,\, \bigg\vert \int_{0}^{1} du \, \frac{u \left( u \cos u - \sin u \right) }{r^4 f(u^2/r^2)} \bigg\vert 
\\
& +  \, c \bigg\vert \int_{1}^{\infty} du \, \frac{u \cos u - \sin u }{u^3} \bigg\vert  ,
\end{split}
\end{equation}
for some constant $c$. 
Since the first integrand on the \textit{r.h.s.} is bounded for $(r,u) \in (0,r_0]\times (0,1]$, the corresponding integral is bounded for $r<r_0$. It is easy to prove that the remaining integral, over an unbounded interval, is finite. 
Therefore, $\lim_{r\to 0}\rho_\text{eff}^\prime(r)$ is bounded even for a non-analytic function $f(k^2)$ which  asymptotically grows as~$k^4$.

The reasoning above proves that if the functions $f(k^2)$ asymptotically grow at least as fast as $k^{4}$, then the potentials $\chi$ are 1-regular, and the curvature invariants in $\mathscr{I}_{0}$ are regular. This comprehends a large class of local and nonlocal theories. In particular, all polynomial-derivative models with more than four derivatives in both scalar and spin-2 sectors have a regular Newtonian limit, and this feature is not changed if the logarithmic quantum corrections are taken into account. Also, the one-loop quantum corrections do not spoil the regularity for the weakly nonlocal models for which $f(k^2)$ tend to a polynomial of the type $k^{n}$ ($n\geqslant 4$) in the UV, as well for the family of models for which the classical $f(k^2)$ is an exponential function~\eq{GFNff}. This result is a generalisation of~\cite{BreTib2} to the case in which the effective sources can be non-analytic functions.

Finally, it is useful to mention that in the case of the fourth-derivative gravity, $f(k^2)\sim k^{2}$ in the UV and $\rho_\text{eff} (r)$ diverges as $r\to 0$, which means that at least one of the conditions in~\eqref{RegCond} are violated. Indeed, the effective source for the classical model was calculated  in~\cite{BreTib2}, and it was shown that it behaves like $r^{-1}$ for small $r$. When the logarithmic one-loop quantum corrections are taken into account,
\beq
f(z) = 1 + z \, [ \al + \be \ln (z/\mu^2)],
\eeq
and the asymptotic behaviour becomes $f(k^2)\sim k^{2}\ln k^2$.
Then, combining the result of Ref.~\cite{Nos4der} with Eq.~\eqref{Lapla} we obtain,
in the small-$r$ approximation,
\begin{equation} \label{Fonte4der}
\rho_\text{eff}(r) \, \underset{r \to 0}{\sim} \,\, \frac{M}{4\pi r} \frac{1}{\al - 2\be \ln(\mu r)},
\end{equation}
which diverges slightly more slowly than in the classical model, but not enough to regularise the potential beyond the 0-regularity~\cite{Nos4der}.

\begin{figure}
\begin{center}
\includegraphics[width=4.0cm,angle=0]{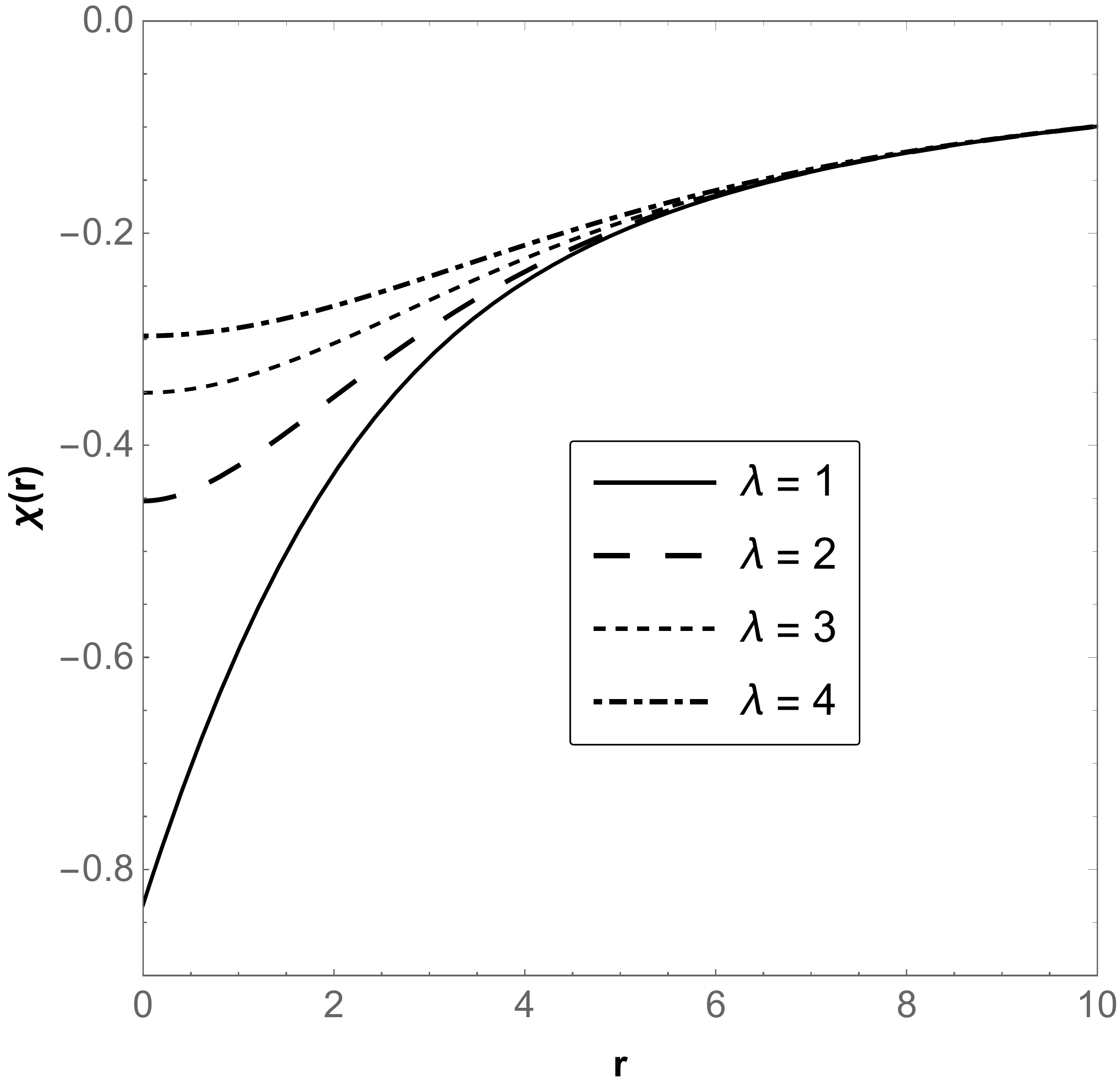}
\includegraphics[width=4.0cm,angle=0]{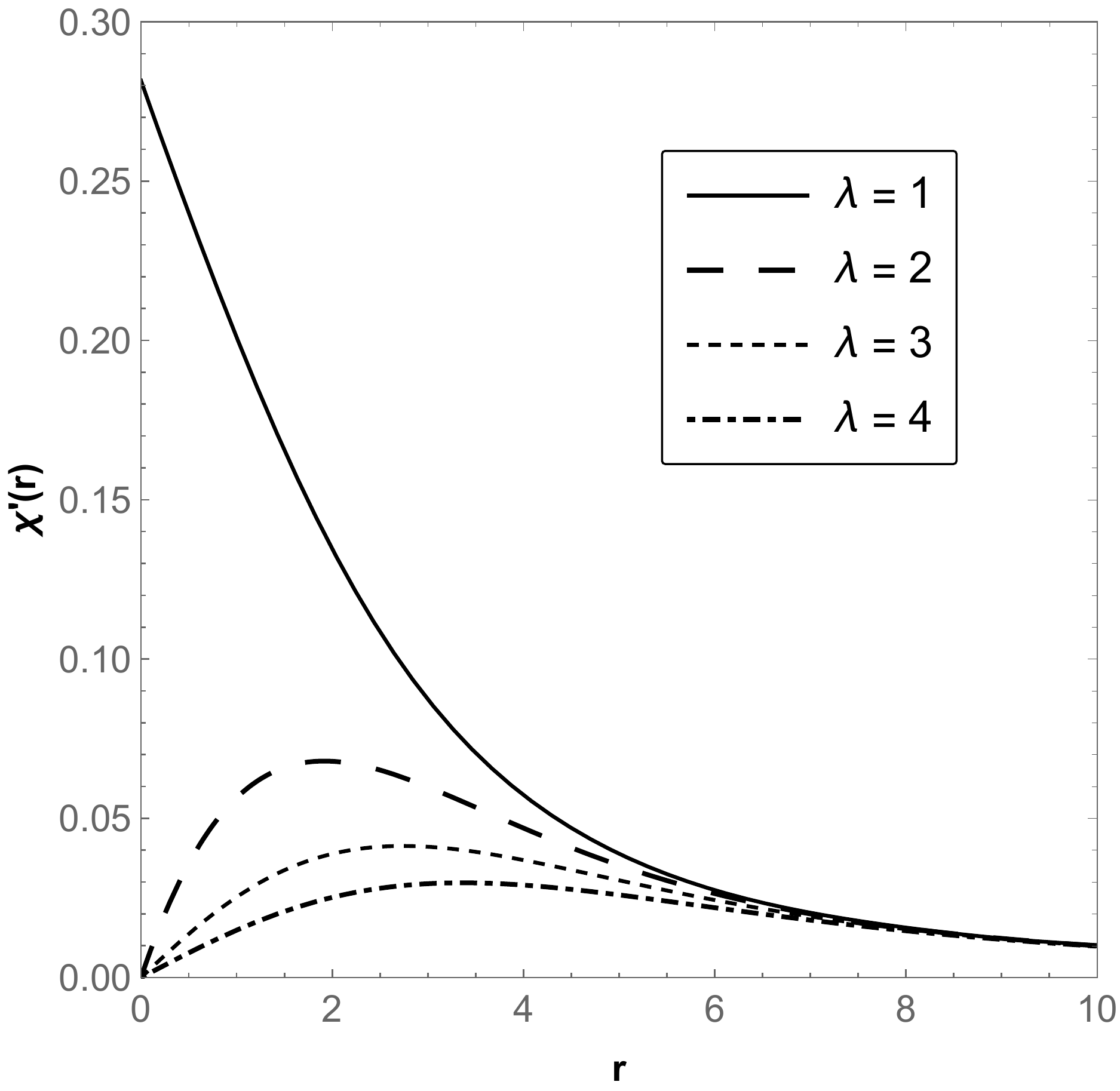}
\\
\includegraphics[width=4.04cm,angle=0]{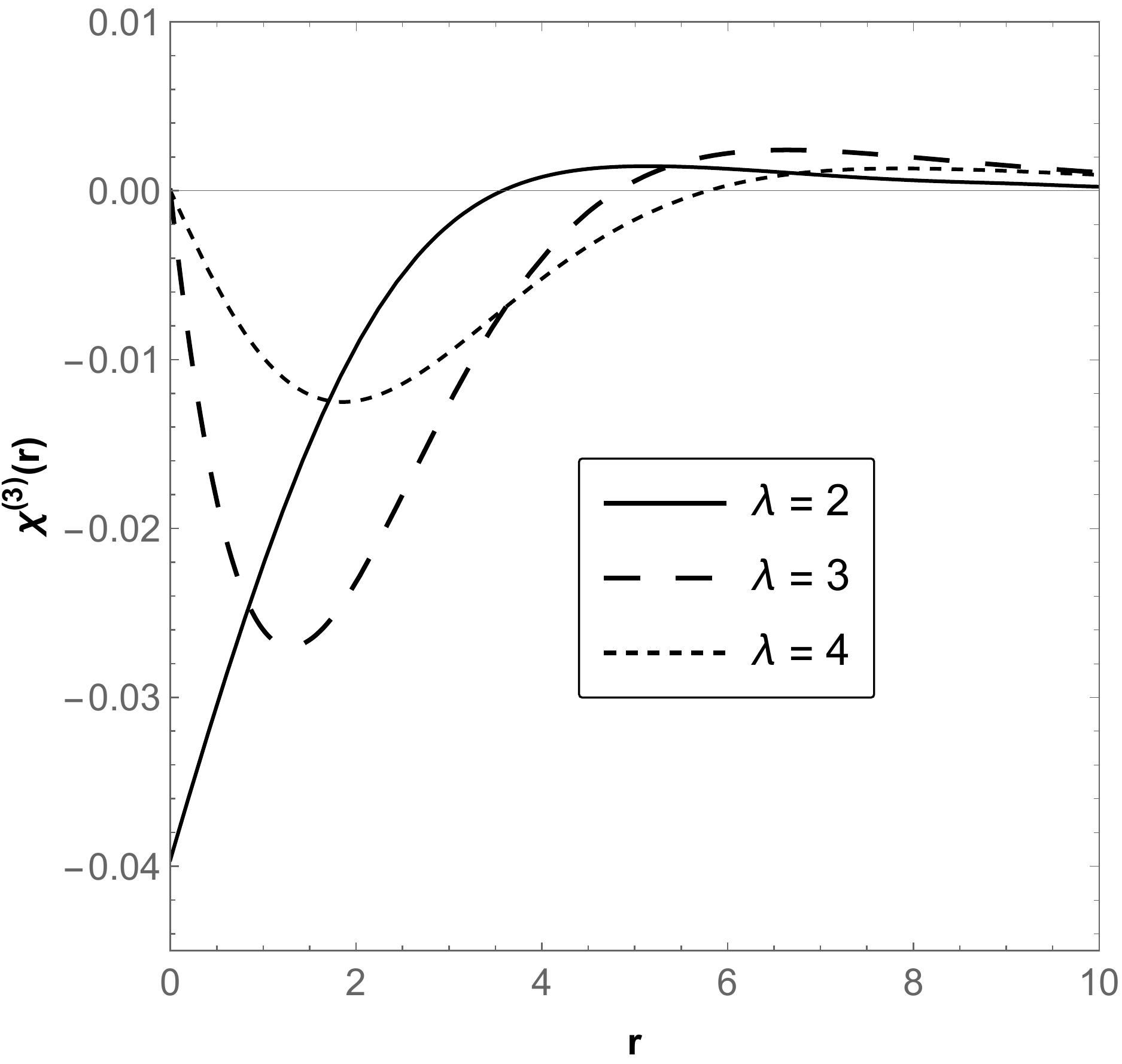}
\includegraphics[width=4.0cm,angle=0]{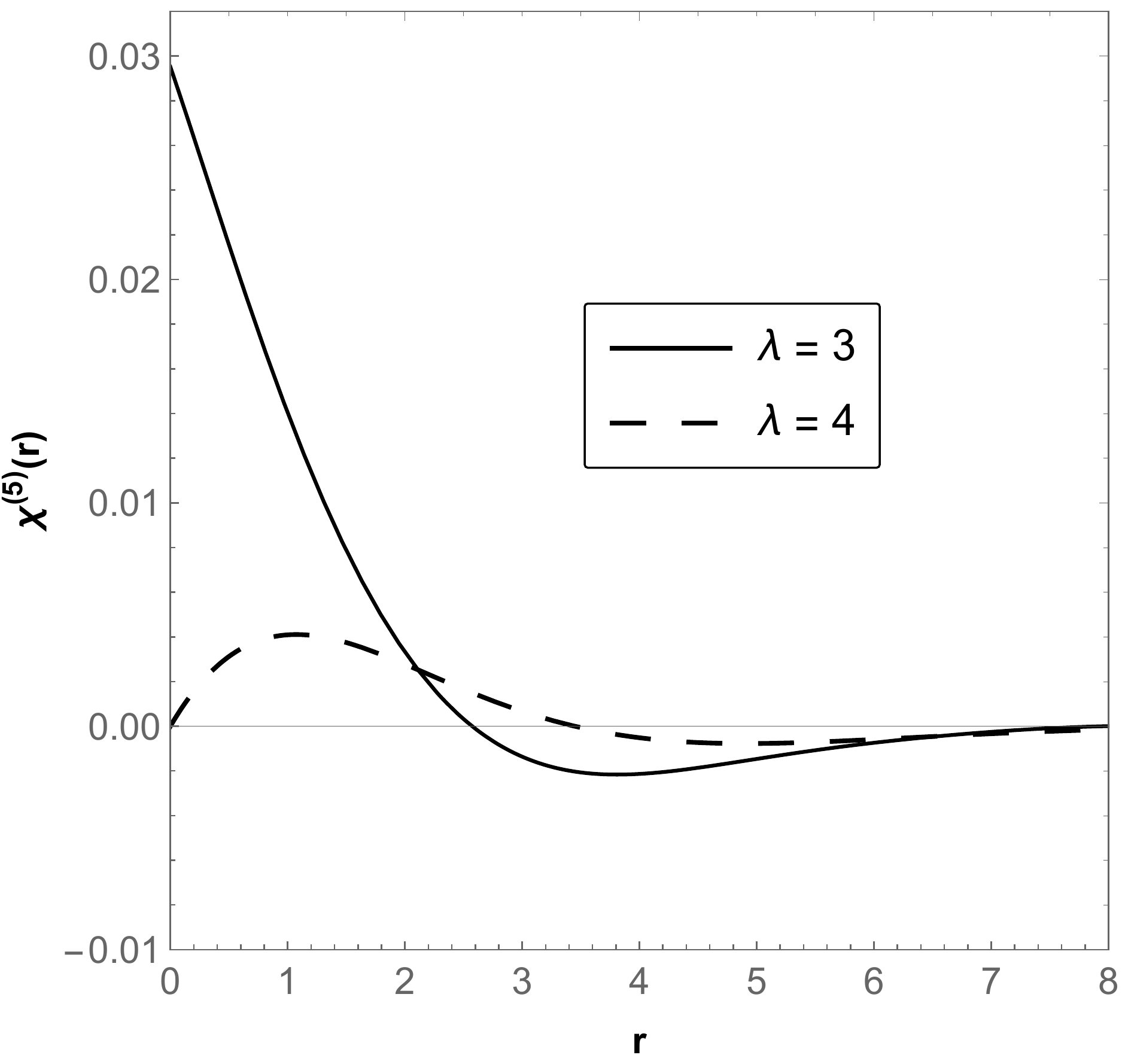}
\caption{ \sl Numerical evaluation of $\chi(r)$ and its first odd-order derivatives for the Kuz'min form factor for $\la\in\lbrace 1,2,3,4\rbrace$ in~\eqref{HKuzCap3}.}
\label{Graph1}
\end{center}
\end{figure}

The effective source also diverges for nonlocal gravity theories defined by the form factors~\eqref{nonlocal} that tend to a constant in the UV. One such example is the Kuz'min form factor~\cite{Kuzmin},
\beq
\label{HKuzCap3}
H(k^2) = \lambda \left[ \ga + \Ga(0,k^2/m^2) + \ln (k^2/m^2) \right]   ,
\eeq
in the case with $\lambda =1$. Here and in the following $\ga$ denotes the Euler-Mascheroni constant, $\Ga(0,z)$ is the incomplete gamma function, $m$ is a mass parameter and $\lambda \in \mathbb{N}$. For large momentum it satisfies 
\beq
\label{HKuzLimCap3}
\lim_{k \rightarrow \infty} \, e^{H(k^2)} \, \approx \, e^{\lambda \ga} \,  \left( \tfrac{k}{m}\right) ^{2 \lambda}  \, ,
\eeq
whence, in the classical theory, $f(k^2) \sim k^2$ if $\lambda = 1$ and the associated potential is only 0-regular. This can be viewed in the numerical evaluation of $\chi(r)$ and $\chi^\prime(r)$ using~\eqref{Chi-Int}, displayed in the upper panels of Fig.~\ref{Graph1}. Notice that for $\la=1$ the potential is finite, but its first derivative does not vanish as $r\to 0$, indicating the singularity of the source (and of the curvature invariants).
As in the local fourth-derivative gravity~\cite{Nos4der}, the leading logarithmic quantum correction does not change this outcome. 
On the other hand, for $\lambda \geqslant 2$ the corresponding potential is at least 1-regular, regardless of the logarithmic corrections, as discussed above.

%%%%%%%%%%%%%%%%%%%%%%%%%%%%%%%%%%%
%%%%%%%%%%%%%%%%%%%%%%%%%%%%%%%%%%%

\subsection{Higher-order regularity}
\label{Sec5b}

The main argument used above to prove the 1-regularity of the potential involved the first derivative of the effective source. It is possible to generalise the previous discussion to higher-order derivatives and investigate the higher-order regularity of the Newtonian potentials.

Accordingly, we first observe that since the function $g(r,k)$ in~\eqref{g_def} is even and analytic in $r$, and it can be expressed as a series,
\begin{equation}
g(r,k) = \frac{k^2}{f(k^2)} \sum_{\ell=0}^\infty \frac{(-1)^\ell}{(2\ell + 1)!}  \left( kr \right)^{2\ell},
\end{equation}
whence,
\beq \label{array0}
\lim_{r \to 0} \, \frac{\partial^{n} }{\partial r^{n}} g(r,k)   =  \left\{ 
\begin{array}{l l}
0 \, ,  &  \text{if } n \text{ is odd},\\
\frac{(-1)^{n/2}}{(n + 1)} \frac{k^{n+2}}{f(k^2)}  \, , &  \text{if } n \text{ is even}.\\
\end{array} \right . \,
\eeq
Furthermore, the derivatives with respect to $r$ are bounded (for a fixed $k$). This can be seen by noticing that since
$$
\frac{\pa^{n+1}}{\pa r^{n+1}} g(r,k) = - \frac{n+1}{r} \frac{\pa^{n}}{\pa r^{n}} g(r,k) + \frac{k^{n+2}}{r f(k^2)} \sin \bigg[ kr + \frac{(n+1)\pi}{2}\bigg] ,
$$
the extrema of $\tfrac{\pa^{n}}{\pa r^{n}} g(r,k)$ are limited by $\tfrac{k^{n+2}}{(n+1)f(k^2)}$. Taking~\eqref{array0} and the analyticity of $g(r,k)$ in $r$ into account, we have
\begin{equation} \label{QuotaDerN}
\bigg\vert \frac{\pa^{n}}{\pa r^{n}} g(r,k) \bigg\vert \leqslant \frac{k^{n+2}}{(n+1)f(k^2)} ,
\end{equation}
which generalises~\eqref{QuotaDer}.

Regarding the upper bound defined by~\eqref{QuotaDerN} as a function of $k$ it follows that if $f(k^2)$ grows at least as fast as  $k^{n+4}$, then the improper integral
$$
\int_0^\infty dk \, \frac{\pa^n}{\pa r^n}g(r,k) 
$$
converges uniformly. As in the previous subsection, under these circumstances the source $\rho_{\text{eff}}$ can be differentiated $n$-times, namely,
\begin{equation} 
\rho_\text{eff}^{(n)}(r) = \frac{M}{2 \pi^2} \int_{0}^{\infty} dk \, \frac{\pa^n}{\pa r^n}g(r,k) 
\end{equation}
and the limit $r \to 0$ can be interchanged with the integral in the expression above. In particular,  the odd derivatives of the source vanish at $r=0$ because of~\eqref{array0}.

This result can be reformulated as: if the function $f(k^2)$ asymptotically grows at least as fast as $k^{4+2N}$ for an integer $N>0$, then the effective source $\rho_\text{eff}(r)$ is (at least) $2N$ times differentiable and $\rho_\text{eff}^{(n)}(0)=0$ for all odd $n \leqslant 2N$---in other words, $\rho_\text{eff}(r)$ is $N$-regular.
As a corollary, we notice that if $f(k^2)$ asymptotically grows faster than any polynomial, then the effective source is an analytic function of $r$ and is $\infty$-regular.

Of course, the converse of the collorary is not true, as in Ref.~\cite{BreTib2} it was shown by explicit calculation that $\rho_\text{eff}(r)$ is also analytic if $f(k^2)$ is a polynomial.
For example, for the polynomial model with $N\geqslant 2$ simple poles considered in Sec.~\ref{class-ex} we have $f(k^2)\sim k^{2N}$ and the analytic function~\cite{BreTib2}
\beq \label{SourcePoly}
\rho_\text{eff}(r) = \frac{M}{4\pi r} \sum_{i=1}^N  C_i \, m_i^2 \, e^{-m_i r} ,
\eeq
with $C_i$ defined in~\eqref{coi}. The finiteness at $r=0$ follows from the result~\eqref{array1} with $n=0$, which holds for any $N\geqslant 2$. The remaining identities can be used to explicitly show that for $N\geqslant 3$ the source~\eqref{SourcePoly} is $(N-2)$-regular.
% (a fact which was not observed in~\cite{BreTib2}).

As this example already suggests, and like in Sec.~\ref{Sec5a}, the higher-order regularity properties of the source can be extended to the Newtonian potential $\chi(r)$. Namely, if the source $\rho_\text{eff}(r)$ is of class $C^{2N}$, the potential is $C^{2N+2}$, and it is straightforward to verify that the $p$-regularity of the source implies that the potential is $(p+1)$-regular. The conclusion is that if the function $f(k^2)$ asymptotically behaves as $k^{4+2N}$ for some $N\geqslant 0$, then the associated potential is $(N+1)$-regular and all the curvature invariants in $\mathscr{I}_{2N}$ are regular.

To close this section, let us return to the example of the classical Kuz'min form factor~\eqref{HKuzCap3}. Since in the UV we have $f(k^2) \sim k^{2\la}$, according to the discussion above the potential $\chi$ must be $(\la-1)$-regular.
In Fig.~\ref{Graph1} we display the numerical evaluation of the first odd-order derivatives of the potential $\chi(r)$ for $\la\in\lbrace 1,2,3,4\rbrace$, which verifies our result and shows that the potential cannot have an order of regularity higher than $\la-1$, just like for the local (polynomial) form factors. Indeed, for a form factor with a certain $\la$, we see that $\chi^{(2\la-1)}(0)\neq 0$. Again, the leading logarithmic quantum corrections do not modify the regularity order of the potential.  A similar consideration applies to the more general nonlocal form factors proposed in Refs.~\cite{Tomboulis,Modesto12}, which also tend to a polynomial in the UV.

%%%%%%%%%%%%%%%%%%%%%%%%%%%%%%%%%%%
%%%%%%%%%%%%%%%%%%%%%%%%%%%%%%%%%%%

\section{Perturbative solution of the potential}
\label{pert}
\label{Sec6}

{\it 
Summary of the section: here we focus on the quantum logarithmic corrections to the Newtonian potentials, treated as the first order correction to the 2-point correlation function. This differs from the approach employed in the previous section, which considered the full resummation of the one-loop 1-particle irreducible dressed propagator. We evaluate the explicit form for the correction at first order in $\be$ for two specific models: the polynomial gravity with simple poles and one ghost-free nonlocal gravity model. General results are also obtained concerning the UV and the IR behaviours of the quantum-corrected potentials.
}\\

Another approach to obtain a solution for the potential with logarithmic quantum corrections is to solve the differential equation~\eqref{EqPot} using perturbation theory in $\beta$. This relies on the assumption that the scale related to the one-loop quantum correction term is much smaller than the classical counterpart because the former is of order $O(\hbar)$. Thence, we shall rewrite Eq.~\eq{efizinho} splitting its classical and quantum parts,
\beq
\n{eq0}
f(z) = f_c (z) +  \be \, z \, \ln (z/ \mu^2) ,
\eeq
and look for solutions of the potentials in the form
\begin{equation} 
\n{eq2}
\chi = \chi_c + \chi_q + O(\beta^2)
,
\end{equation}
where $\chi_c$ is the $O(\be^0)$ classical potential and  $\chi_q$ is the  $O(\be)$ one-loop correction. 

Substituting \eq{eq0} and \eq{eq2} into \eq{EqPot} gives the equations at each order in $\beta$,
\begin{align}
f_c (-\De) \De \chi_c &= \ka M \de ( \vec{ r} )
,
\label{Beta0}
\\
f_c (-\De) \De \chi_q &= \be \ln (-\De/ \mu^2) \De^2 \chi_c
.
\label{Beta1}
\end{align}
Hence, in the three-dimensional Fourier space we have the transformed potentials
\begin{align}
\tilde{\chi}_c (k) &= - \frac { \ka M } { k^2 f_c (k^2) }
,
\\
\tilde{\chi}_q (k) &= - \frac{\be k^2} {f_c (k^2)} \,  \ln \left( \frac{k^2}{\mu^2} \right)  \tilde{\chi}_c (k) ,
\end{align}
which after the integration over the angular coordinates yield
\begin{equation} 
\n{class-po2}
\chi_c (r) = - \frac{\ka M}{2 \pi ^2 r} \int_0^\infty dk\, \frac{\sin (kr)}{k f_c (k^2)}
\end{equation}
and
\begin{equation} 
\n{quan-po}
\chi_q (r) =  \frac{\be \ka M}{ \pi ^2 r} \int_0^\infty dk  \,\frac{k \sin (kr)  \ln (k/\mu)}{\left[ f_c (k^2)\right]^2}\, 
.
\end{equation}

\begin{figure}
\begin{center}
\includegraphics[width=6.5cm,angle=0]{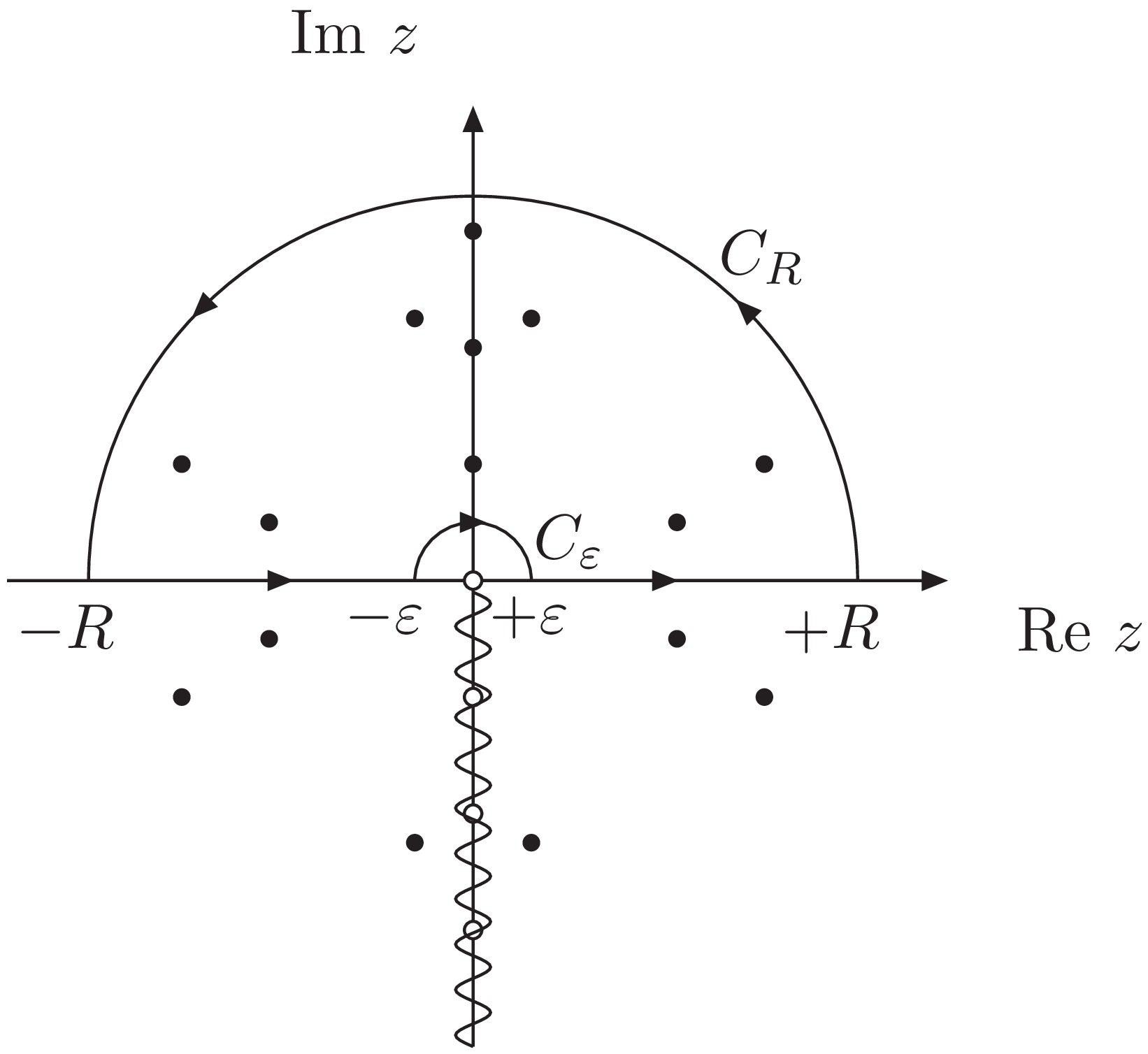}
\caption{\sl Contour of integration used to evaluate~\eqref{eq3}, poles and
branch cut defined by~\eqref{eqt1}.}
\label{con}
\end{center}
\end{figure}

As mentioned above, the potential $\chi_c$ given by \eq{class-po2} coincides with the one without logarithm quantum corrections, whose explicit solution for different types of classical gravity models can be found, \textit{e.g.}, in~\cite{Stelle77,Modesto12,Newton-MNS,Newton-BLG,BreTib1,Head-On,Edholm:2016hbt,Tseytlin:1995uq,Accioly:2016qeb}. In particular, the case of polynomial-derivative gravity with simple poles has been discussed in detail in Sec.~\ref{class-ex}. Therefore, our main concern here involves the integral
\beq \label{W_dpn}
I \equiv \int_0^\infty dx  \,\frac{x \sin (xr) \ln(x/\mu)}{\left[ f_c (x^2)\right]^2}  ,
\eeq
which appears in Eq.~\eq{quan-po}. In what follows, we shall evaluate it explicitly for the two models discussed in Sec.~\ref{class-ex}: the local polynomial gravity and the exponential ghost-free gravity. Subsequently, we present general results regarding the regularity of $\chi_q$ and its behaviour in the IR regime.

%%%%%%%%%%%%%%%%%%%%%%%%%%%%%%%%%%%
%%%%%%%%%%%%%%%%%%%%%%%%%%%%%%%%%%%

\subsection{Polynomial gravity}
\label{pert_A}

%%%%%%%%%%%%%%%%%%%%%%%%%%%%%%%%%%%
%%%%%%%%%%%%%%%%%%%%%%%%%%%%%%%%%%%

The first example we consider is the case in which $f_c$ is a polynomial function with only simple roots, given by Eq.~\eq{fz-poly} with $N>1$. 
Differently from the general formula~\eq{Chi-Int}, in~\eqref{W_dpn} the logarithm function appears in the numerator because of the perturbative expansion. Hence, we can follow the method developed in~\cite{Nos4der} and apply Cauchy's residue theorem in the context of the pole structure of the classical theory. 

Let us define the function
\beq
\n{eqt1}
h(z) = \frac{z e^{irz} \ln(z/\mu)}{\left[ f_c (z^2)\right] ^2}, 
\quad - \frac{\pi}{2} \leqslant \,\text{arg}\, z < \frac{3\pi}{2}
\eeq
which has poles at $z = \pm i m_i$ $(i = 1,\cdots,N)$. 
The branch cut defined in \eq{eqt1} corresponds to the negative part of the imaginary axis, therefore it is possible to construct the oriented simple closed
path $C$ depicted in~Fig.~\ref{con}, for which $\Im (z) \geqslant 0$. Notice that $C$ has an indentation around $z=0$, where $\ln z$ is not defined. Since there is only a finite number of poles, we can take $R > \max_i\lbrace |m_i| \rbrace$ and $\vp < \min_i \lbrace |m_i| \rbrace$. Only the poles at $z=+ i m_i$ are inside $C$, then
\beq
\n{eq3}
\ointctrclockwise_C dz\, h(z) = 2 \pi i \sum_{i} \Res (h(z),i m_i).
\eeq
On the other hand,
\begin{equation}
\begin{split}
\n{eq4}
&
\ointctrclockwise_C dz \, h (z) =  \int_\varepsilon^R dx \, \frac{x e^{irx} \ln (x/\mu)}{\left[ f_c (x^2)\right] ^2} 
+ \int_{C_R} dz \, h(z)
\\
& \hspace{2.0 mm}
+ \int_{-R}^{-\varepsilon} dx \, \frac{x e^{irx} (\ln (|x|/\mu) + i \pi)}{\left[ f_c (x^2)\right] ^2} 
+ \int_{C_\varepsilon} dz \, h(z)
.
\end{split}
\end{equation}
Utilizing Jordan's lemma, it follows that the integral along the semicircular arc $C_R$ vanishes when $R \to \infty$; similar consideration shows that the integral along $C_\vp$ also vanish in the limit $\vp \to 0$. Thus, making the substitution $x \mapsto - x$ in the third integral in the {\it r.h.s.} of formula~\eq{eq4} and comparing with~\eq{eq3} one has 
\beq
\n{K-an}
I
= \pi \Re \Big[ \sum_{i} {\Res (h(z), i m_i) } \Big] +  \frac{\pi}{2} \int_0^\infty dx \, \frac{x \cos (rx)}{\left[ f_c (x^2)\right] ^2} 
.
\eeq

In order to evaluate the remaining integral in~\eq{K-an}, we employ the partial fraction decomposition, which now gives (\textit{c.f.}~\eqref{PartFrac1})
\begin{equation}
\n{fp-dec}
\frac{1}{\left[ f_c (z^2)\right] ^2} = \sum_{i=1}^N \sum_{j=1}^2 A_{i,j} \, \left( \frac{m_i^2}{z^2+m_i^2} \right)^j
,
\end{equation}
with the coefficients
\begin{equation}
A_{i,1} = 2 m_i^2 \sum_{k \neq i} \frac{ m_k^4}{(m_i^2 - m_k^2)^3} \prod_{\ell \neq i, k} \left( \frac{m_\ell^2}{m_i^2 - m_\ell^2} \right)^2
\end{equation}
and
\begin{equation}
A_{i,2} =  \prod_{j \neq i} \left( \frac{m_j^2}{m_j^2 - m_i^2} \right)^2.
\end{equation}
Therefore, using~\eq{fp-dec} we get 
\beq \label{617}
\begin{split}
& \hspace{-1mm}
\int_0^\infty dx \, \frac{x \cos (rx)}{\left[ f_c (x^2) \right] ^2}
=  \sum_{i=1}^N \left[ m_i^4\, A_{i,2} \int_0^\infty dx \, \frac{x \cos (rx)}{(x^2+m_i^2)^2} 
\right. \\
& \hspace{2cm} \left. + \, m_i^2\, A_{i,1} \int_0^\infty dx \, \frac{x \cos (rx)}{x^2+m_i^2}
\right].
\end{split}
\eeq
%%%%
The above integrals have the result (for $\Re (m_i^2) \geqslant 0$)~\cite{book2} 
\beq
\begin{split}
\n{Raabe}
&
\int_0^\infty dx \, \frac{2 m_i^2 x \cos (rx)}{(x^2+m_i^2)^2} = 1
+ m_i r  \sinh (m_i r) \,\text{Chi}(m_i r)
\\
& \hspace{0.45cm}
- m_i r \cosh (m_i r) \,\text{Shi}(m_i r)
,
\end{split}
\eeq
and
\beq
\begin{split}
\n{Raabe2}
&
\int_0^\infty dx \, \frac{x \cos (rx)}{x^2+m_i^2} =  \sinh (m_ir) \, \text{Shi}(m_ir)
\\
& \hspace{2cm}
- \cosh (m_ir) \, \text{Chi}(m_ir)
,
\end{split}
\eeq
%\\
where we define the hyperbolic integrals
\begin{align}
&\mbox{Chi} (z) =  \ga + \ln z + \int_0^z dt\, \frac{\cosh t - 1}{t} 
,
\\
&\mbox{Shi} (z) = \int_0^z dt\, \frac{\sinh t}{t}.
\end{align}

\begin{widetext}
On the other hand, the real part of the residues in \eq{K-an} is given by
\beq
\begin{split}
\label{622}
\Re \Big[ \sum_{i} {\Res (h(z), i m_i) } \Big] =
\sum_{i=1}^N \frac{ m_i^2 e^{-m_i r}}{4}
\left[ 2 A_{i,1} \ln \Big (\frac{ m_i}{\mu } \Big)
+ A_{i,2} m_i r \ln \Big( \frac{m_i}{\mu } \Big) - A_{i,2}  \right].
\end{split}
\eeq
Collecting~\eqref{617}, \eqref{Raabe}, \eqref{Raabe2} and~\eqref{622} we get the expression for the quantum correction to the potential,
\beq
\begin{split}
\n{loop-pot}
&
\chi_q(r) =  \frac{\be \ka M}{4\pi r } \sum_{i=1}^N m_i^2
\left\lbrace 
A_{i,2} (1 - e^{-m_i r} ) + (2 A_{i,1} + A_{i,2} m_i r ) \, e^{-m_i r} \ln (m_i/\mu)   
\right.
\\
&
\left.
\hspace{4mm}
+ \, \big[ 2   A_{i,1} \sinh (m_ir) - A_{i,2} m_i r \cosh(m_i r) \big] \, \text{Shi}(m_ir)
- \big[ 2 A_{i,1} \cosh (m_i r) - A_{i,2} m_i r \sinh (m_ir) \big] \, \text{Chi}(m_ir)
\right\rbrace .
\end{split}
\eeq

Taking into account the series representations of the functions above, the potential \eq{loop-pot} can be written as
\beq \label{ChiQ-Series}
\chi_q(r) =  \frac{\be \ka M}{4\pi} \sum_{k=0}^{\infty} \left\lbrace \frac{2}{(2k)!}\Big[H_{2k}-\gamma-\log(\mu r) \Big] \, G_{k} \, r^{2k-1}+ \frac{1}{(2k+1)!} E_{k} \, r^{2k} \right\rbrace
\eeq
\end{widetext}
where
\beq
H_k = \sum_{n=1}^k \frac{1}{n}
\eeq
is the $k$-th harmonic number, and
\begin{align}
\
G_{k} =& \sum_{i=1}^N m_i^{2(k+1)}\Big(A_{i,1}-k A_{i,2} \Big) 
,
\\
E_{k} =& \sum_{i=1}^N  m_i^{2k+3} \left\lbrace A_{i,2}-\log\left( \tfrac{m_i}{\mu}\right) \big[2A_{i,1}- (2k+1)A_{i,2} \big]\right\rbrace .
\end{align}
%%%%%%
With a procedure similar to the one of Sec.~\ref{class-ex} one can prove that
\beq \label{Gi-rel}
k \leqslant 2N-2 \quad \Longrightarrow \quad G_{k} = 0,
\eeq
whereas
\beq
G_{2N-1} = - \prod_{i=1}^N m_i^4 \neq 0.
\eeq
%%%%%%
Due to these relations, the non-analytic part of \eq{loop-pot} has the general structure 
\begin{equation} \label{LogStruc}
r^{2k-1} \ln (\mu r) 
\quad \text{with} \quad
k \geqslant 2N-1.
\end{equation}
Moreover, the potential $\chi_q$ is $(2N-2)$-regular and the one-loop quantum correction does not spoil the regularity of the classical solution. In fact, in Sec.~\ref{class-ex} we proved that $\chi_c$ is $(N-1)$-regular.
The explicit calculations of this section are also in total agreement with the general discussion of Sec.~\ref{source-gen}; we shall return to this issue in Sec.~\ref{pert_C}.

Finally, we point out that even though the potential $\chi_q$ associated to the fourth-derivative gravity~\cite{Nos4der} can be obtained from the general expression~\eq{loop-pot} by setting $N=1$, in such a case relations~\eq{Gi-rel} only hold for $G_0$. In that case
we have $A_{1,1} = 0$ and $A_{1,2} = 1$, so that $\chi_q(r)$ is finite (0-regular) but does not satisfy the $1$-regularity conditions~\eqref{RegCond} as the leading non-analytic term is already $r\ln (\mu r)$ (see~\cite{Nos4der} for further discussion on this model).

%%%%%%%%%%%%%%%%%%%%%%%%%%%%%%%%%%%
%%%%%%%%%%%%%%%%%%%%%%%%%%%%%%%%%%%

\subsection{Nonlocal ghost-free gravity}
\label{pert_B}

As a further example of a form factor that admits a compact expression for the quantum correction $\chi_q (r)$, let us
consider the simplest case of nonlocal ghost-free gravity, for which 
\beq
f_c (k^2) = e^{k^2/m^2}. 
\eeq
In this case, the integral \eq{W_dpn} is given by 
\beq \label{W_gf}
I = \int_0^\infty dx  \,x e^{-2x^2/m^2} \, \sin (xr) \ln(x/\mu),
\eeq
whose solution can be obtained by taking a derivative (with respect to $a$) of the integral representation of Kummer's function~\cite{Grad,NIST},
$$
M \left(  \tfrac{3}{2} -a, \tfrac{3}{2},-\tfrac{z^2}{4b} \right)  =   \frac{2 b^{\tfrac32 - a}}{z \, \Ga \left(\tfrac32 -a\right)} \int_0^\infty dx \, 
 x^{1 - 2a} e^{-bx^2} \sin (z x)
,
$$
defined for $\Re (a) < 3/2, \Re(b) >0$, and using the property 
$
M(a,b,z) = e^z M(b-a,b,-z)
$.
The final result for~\eq{quan-po} reads
\begin{equation}
\begin{split} \label{Chi1Exp}
\chi_q(r) = & \,\, \frac{\beta \kappa M m^3}{8 (2 \pi)^{3/2}} e^{-\frac{m^2r^2}{8}} \bigg[ 2 - \gamma +  2\log \bigg( \frac{m}{8 \mu} \bigg) 
\\
& \,\, - \frac{\partial}{\partial a} \, M\left(a,\tfrac{3}{2},\tfrac{m^2 r^2}{8}\right) \Big\vert_{a=0} \bigg].
\end{split}
\end{equation}

The solution above is analytic because Kummer's confluent hypergeometric function,
\beq
M(a,b,z) = {}_1 F_1(a;b;z) = \sum_{n = 0}^\infty \frac{(a)_n z^n}{(b)_n n!},
\eeq 
is entire for $b=3/2$. Here,
$(p)_{n}=p(p+1)\cdots(p+n-1)$ is the Pochhammer symbol.
The last term inside the brackets in \eq{Chi1Exp}, actually, has a simple power series representation: from $(a)_n = a (n-1)! + O(a^2)$, we get     
$ \lim_{a \to 0} \pa_a (a)_n = (n-1)! \,$, so that
\begin{equation} 
\frac{\partial}{\partial a} \, M\left(a,b,x \right)\Big\vert_{a=0} = \sum _{n=1}^{\infty } \frac{  x^n}{n \left( b\right) _n} \, .
\end{equation}
Therefore, since the functional dependence of Eq.~\eq{Chi1Exp} involves only $r^2$, the quantum correction to the potential is $\infty$-regular (see discussion in Sec.~\ref{Sec3}), just like its classical counterpart, given by Eq.~\eq{Tseytlin}.
Finally, the analyticity of the potential $\chi_q$ in this example can be qualitatively explained as the limiting scenario of the polynomial gravity (discussed in Sec.~\ref{pert_A}) when the number of derivatives in the action tends to infinity; thence, the logarithmic terms which occur in~\eqref{LogStruc} hide in the infinity.

%%%%%%%%%%%%%%%%%%%%%%%%%%%%%%%%%%%
%%%%%%%%%%%%%%%%%%%%%%%%%%%%%%%%%%%

\subsection{Regularity in general higher-derivative gravity}
\label{pert_C}

The procedure to evaluate $\chi_q$ can be cumbersome for more general higher-derivative models defined by other nonlocal form factors, or by polynomial functions $f_c$ which contain degenerate roots. However, some general properties of the one-loop quantum correction to the potential in these models can be derived without the need to work out the explicit solution. 

In what concerns the small-$r$ behaviour, this follows from the observation that $\chi_q$ is sourced by a smeared distribution, see Eq.~\eqref{Beta1}. In fact, its integral representation~\eqref{quan-po} is very similar to the one of the source~\eqref{rho_def}, both integrands being regular as $k \to 0$.
Therefore, we can apply the same formalism of Sec.~\ref{source-gen} to the potential $\chi_q$, which is the solution of
\begin{equation} 
\De \chi_q = \ka \, \be \, \tilde{\rho}_{\text{eff}} \, ,
\end{equation}
where
\begin{equation} \label{RhoP1}
\tilde{\rho}_{\text{eff}}(r) = \frac{M}{2 \pi ^2} \int_0^\infty dk  \,\frac{k^3 \sin (kr) \ln (k^2/\mu^2) }{r \left[ f_c (k^2)\right]^2}  
\, .
\end{equation}
Rewriting the integrand of~\eqref{RhoP1}  as (\textit{c.f.}~\eqref{rho_def})
\begin{equation} \label{104}
 g_q(r,k) = \frac{k \sin(kr)}{r \, \phi(k^2)} \quad \text{with} \quad \phi(k^2) = \frac{\big[ f_c (k^2)\big]^2}{k^2 \ln \left(  k^2/ \mu^2\right)} \, ,
\end{equation}
one can follow all the discussion of Sec.~\ref{source-gen} concerning the finiteness and higher-order regularity of the potentials by analysing the function $\phi(k^2)$ instead of $f(k^2)$. Notice that even though $\phi(k^2)$ diverges as $k \to 0$ (while in Sec.~\ref{source-gen} it is assumed that $\lim_{k\to 0}f(k^2) = 1$), both integrands $g_q$ and $g$ vanish as $k \to 0$. In this sense, for most of the cases, here the behaviour for small $k$ is actually improved with respect to the one of Sec.~\ref{source-gen}. Regarding the behaviour for large $k$, if $f_c (k^2) \sim k^{2n}$ asymptotically, then $\phi(k^2) \sim k^{2(2n-1)}/\log k$. This shows that for $n>1$ it happens that $\phi$ grows faster\footnote{On the other hand, for $n=1$ (as in the fourth-derivative gravity) the situation is the opposite---see~\cite{Nos4der} for further discussion on this particular model. Still, the quantum correction $\chi_q$ is finite at $r=0$, like the classical potential.} than $f_c$.

The analysis of Sec.~\ref{source-gen}, \textit{mutatis mutandis}, allows us to conclude that $\chi_q$ is finite (0-regular) if the classical action contains at least four derivatives of the metric---or, in the case of nonlocal theories, if the associated $f_c (k^2)$ asymptotically grows at least as fast as $k^2$. Moreover, if $\chi_c$ is $p$-regular, then $\chi_q$ is $(2p)$-regular. The explicit examples involving the polynomial models considered above perfectly agree with this general result. In short, for the higher-derivative models considered in this work, the perturbative quantum correction to the potential is at least as regular as the classical part, and in most of the cases it is regular at a higher order.

%%%%%%%%%%%%%%%%%%%%%%%%%%%%%%%%%%%
%%%%%%%%%%%%%%%%%%%%%%%%%%%%%%%%%%%

\subsection{Infrared limit}
\label{pert_D}

Regarding the far-IR limit, it is not difficult to see that the large-$r$ leading quantum corrections to the classical mechanics'  Newtonian potential are captured by the $O(\beta)$ correction \eq{quan-po}. This can be verified by using \eq{eq0} into \eq{Chi-Int} and making the change of integration variable $kr \mapsto u$, which yields
\beq
\chi(r) = -\frac{\ka M}{2 \pi^2 r }\int_0^\infty du \frac{\sin u}{u \Big[ f_c(u^2/r^2) + \tfrac{2 \be u^2}{r^2} \ln (u/\mu r) \Big] }.
\eeq
Therefore, for large $r$,
\beq
\begin{split}
\n{nova}
\chi(r) \underset{r \to \infty}{\sim} &- \frac{\ka M}{2 \pi^2 r }\int_0^\infty {du} \, \frac{\sin u}{u f_c(u^2/r^2)}
\\
& + \frac{\ka M \beta}{\pi^2 r^3 } \int_0^\infty du \, \frac{u \sin u \ln (u/\mu r)}{f_c(u^2/r^2)}
.
\end{split}
\eeq
The last integral in \eq{nova} is just \eq{quan-po} in the new variables. Noticing that $f_c(u^2/r^2) \rightarrow 1$ for large values of $r$ (see Eq.~\eqref{efizinho}), the first term in \eq{nova} gives the  classical Newton potential,
\beq
\chi_c(r) \underset{r \to \infty}{\sim} - \frac{\ka M}{4 \pi r },
\eeq
while the quantum part tends to
\begin{equation}
\begin{split}
\chi_q (r) \underset{r \to \infty}{\sim} & \,\, \frac{\be \ka M}{ \pi ^2 r^3} \Bigg[  \int_0^\infty du  \, u \sin u \log u
\\
& \,\, - \log (\mu r) \int_0^\infty du  \, u \sin u  \Bigg] .
\end{split}
\end{equation}
Using the exponential regularization to the distributional integrals above, the result of the first integral is $-\pi/2$ and the other one, a $\delta$-function, from which we obtain
\begin{equation} \label{limitIR}
\chi_q (r) \underset{r \to \infty}{\sim } - \frac{\be \ka M}{ 2 \pi r^3} ,
\end{equation}
for \textit{any} classical higher-derivative model defined by analytic form factors.
This result is in agreement with the common lore about the effective theory of quantum gravity, which states that the details of the underlying ``true'' quantum theory of gravity are unimportant for the behaviour of the low-energy regime~\cite{Duff:1974ud,Donoghue:1993&94,Muzinich:1995uj,Hamber:1995cq,Dalvit:1997yc,Akhundov:1996jd,Khriplovich:2002bt,BjerrumBohr:2002kt}. 

Of course, the particular cases presented in the previous subsections exemplify this general result. Indeed, by means of the explicit solutions obtained, Eqs.~\eq{loop-pot} and~\eqref{Chi1Exp}, the large-$r$ limit~\eqref{limitIR} of the quantum corrections to the potential can be directly verified.

%%%%%%%%%%%%%%%%%%%%%%%%%%%%%%%%%%%
%%%%%%%%%%%%%%%%%%%%%%%%%%%%%%%%%%%

\section{Summary and conclusion}
\label{Sec7}

The present work can be regarded as a generalisation of previous results concerning the possibility of avoiding spacetime singularities in higher-derivative theories of gravity (see, \textit{e.g.},~\cite{BreTib1,BreTib2,Newton-MNS,Newton-BLG,Frolov:Poly,Frolov:Exp,Head-On,Edholm:2016hbt,Accioly:2016qeb,Zhang14,Modesto-LWBH}). Due to the difficulties in obtaining exact solutions for the full non-linear theory, most of the results in the literature are derived in the Newtonian limit. Here, the considerations were still restricted to the linearized version of the model, but we made two generalisations.

Instead of solely considering the curvature invariants made only by curvature tensors, we also discussed the regularity properties of scalars containing derivatives of the curvatures. In this vein, the main result was
a relation between the number of derivatives in the action and the maximum number of derivatives in the regular scalars: 
 all the curvature-derivative invariants with at most $2n$ derivatives of curvatures are regular if the local gravity action has at least $2n+6$ derivatives in both scalar and spin-2 sectors (moreover, there are scalars with $2n+2$ derivatives which are singular). The regularity of \textit{all} the local curvature invariants can be achieved in some classes of nonlocal gravity, namely, those defined by a form factor that grows faster than any polynomial, in the~UV.
So far, the known solutions that are ``infinitely regular'' are the Nicolini--Smailagic--Spallucci metric 
\cite{Nicolini:2008aj, Nicolini:2005vd} and similar generalisations~\cite{Zhang14}.
Other known solutions with singularity-free Kretschmann invariant may have higher-order divergences. One example is  Dymnikova's metric~\cite{Dymnikova:1992ux}, for which a direct evaluation of $\Box^2 R$ reveals a divergence at $r = 0$. This is in accordance
with the analysis of the present paper since, in that solution, the Taylor expansion of the metric components has the first non-zero odd-order coefficient at $O(r^5)$.

Furthermore, in our analysis we also allowed for the possibility of some  universal non-analytic form factors associated with quantum corrections. The conclusion is that the {\it logarithmic corrections} do not change the regularity of the Newtonian limit, inasmuch as they are sub-leading with respect to the classical part of the form factor in all the super-renormalizable models.

Since the main set of theories considered in this paper have a classical propagator at least as strong as $k^{-6}$,  the quantum-correction $k^4 \log k^2$ has a more prominent role in the far-IR regime, where it gives the leading correction to Newton's potential, proportional to $ \beta r^{-3}$. We showed that this qualitative behaviour is not affected by the specific classical action (the values of the quantities $\be$ are model-dependent, though). This result supports the hypothesis of the universality of the effective approach to quantum gravity in the IR~\cite{Duff:1974ud,Donoghue:1993&94,Muzinich:1995uj,Hamber:1995cq,Dalvit:1997yc,Akhundov:1996jd,Khriplovich:2002bt,BjerrumBohr:2002kt}. Also, it is worth mentioning that we evaluated the quantum correction to the potential to linear order in $\be$ for two models, \textit{viz.} the polynomial-derivative gravity with simple poles in the propagator and one case of nonlocal ghost-free gravity.
Such computations can be viewed as related to the first-order correction to the 2-point correlation function.

Last but not least, we would like to stress the correctness of our result in the ultraviolet regime, regardless the linear approximation. Indeed, the asymptotic freedom of the theory~\cite{Briscese:2019twl} at short distances guarantees the stability of the potential (or the metric) under nonlinear corrections.

%%%%%%%%%%%%%%%%%%%%%%%%%%%%%%%%%%%
%%%%%%%%%%%%%%%%%%%%%%%%%%%%%%%%%%%

%%%%%%%%%%%%%%%%%%%%%%%%%%%%%%%
\begin{acknowledgements}
This work was supported by the Basic Research Program of the Science, Technology and Innovation Commission of Shenzhen Municipality (grant no. JCYJ20180302174206969).
\end{acknowledgements}

%%%%%%%%%%%%%%%%%%%%%%%%%%%%%%%%%%%
%%%%%%%%%%%%%%%%%%%%%%%%%%%%%%%%%%%

\appendix
\section{Proof of the Theorem of Sec.~\ref{Sec3}}
\label{APP}

Here we present the main steps of the proof of the theorem stated in Sec.~\ref{Sec3}, namely,
that given an $n\geqslant 0$ the sufficient condition for the regularity of all the elements in $\mathscr{I}_{2n}$ is that the potentials $\chi_{0,2}$ are $(n+1)$-regular. We choose to work in the isotropic coordinate system, with spherical coordinates $(r,\th,\phi)$ for the spatial sector, \textit{i.e.}, the flat-space metric reads
\beq \label{met}
\eta_{tt}=-1 , \quad \eta_{rr}=1, \quad \eta_{\th\th}=r^2 , \quad \eta_{\phi\phi}=r^2 \sin^2 \th .
\eeq
Thus, the non-zero components of the Riemann tensor associated with the metric perturbation in~\eqref{Metric} are
\begin{align} \label{Riem}
& R_{trtr} = -\ph'' 
\,, \qquad 
R_{t\th t\th} = - r \, \ph' = \frac{
{  R_{t\phi t\phi}   }
}{\sin^2 \theta}
\,,
\\
& R_{r\th r \th} = r \, (\psi ' + r \, \psi '' ) = \frac{
{  R_{r\phi r \phi}   }
}{\sin^2 \theta}
\,,
\quad 
R_{\th\phi\th\phi} =
2 r^3 \psi ' \sin^2 \theta  
\,.
\nonumber
\end{align}
Since $R_{\al\be\ga\de}$ is already of order $\varkappa^2$, in the linear approximation all the covariant derivatives in terms like $\na_{\mu_1} \na_{\mu_2} \cdots \na_{\mu_i} R_{\al\be\ga\de}$ are evaluated in flat spacetime (thus they commute), with the nonzero Christoffel symbols
\begin{equation} \label{Chris}
\begin{split}
&\Ga^r_{\th\th} = - r , \quad \Ga^r_{\phi\phi} = - r \sin^2 \th , \quad \Ga^\th_{\phi\phi} = - \cos\th \sin\th , 
\\
&\Ga^\phi_{\phi\th} = \cot\th , \quad \Ga^\phi_{\phi r} = \Ga^\th_{\th r} = r^{-1} .
\end{split}
\end{equation}
With these ingredients we can evaluate any curvature-derivative scalar $S_{2n}^q \in \mathscr{I}_{2n}$ involving $2n$ derivatives and $q$ curvature tensors. The building blocks of such scalars will always have the structure
\beq \label{BB}
\na_{\mu_1}\na_{\mu_2}\cdots\na_{\mu_i} R_{\al\be\ga\de} , \qquad i=0,1,\cdots, 2n.
\eeq

Since the components of the metric~\eqref{met} do not have the same dimension, let us define the \textit{balanced component} of a covariant rank-$\ell$ tensor $\mathbf{T}$ in isotropic spherical coordinates, denoted by $]T_{\mu_1,\cdots,\mu_\ell}[$, as follows: if  $T_{\mu_1,\cdots,\mu_\ell}$ is a component of $\mathbf{T}$ such that $s$ indices ($0 \leqslant s \leqslant \ell$) are angular indices ($\th$ or $\phi$), then ${]T_{\mu_1,\cdots,\mu_\ell}[} \equiv r^{-s} \, T_{\mu_1,\cdots,\mu_\ell}$. This definition is motivated from the fact that in an invariant $S_{2n}^q$ all the indices of the building blocks~\eqref{BB} are contracted (possibly with other building block) and the contraction of a pair of angular indices involves $\eta^{\th\th}$ or $\eta^{\phi\phi}$, which are proportional to $r^{-2}$. Therefore, in the balanced component we distribute the factor $r^{-2}$ coming from the angular components of the inverse metric between its two indices.

It is also useful to recall the definition of $p$-regularity (introduced in Sec.~\ref{Sec3}) of a function $\pi(r)$, which is of class $C^{2N}$ (with $N\geqslant p \geqslant 0$) and its first $p$ odd-order derivatives vanish as $r\to 0$; we shall denote this property symbolically as $\mathscr{P}(\pi)=p$. 
 According to Taylor's theorem, if $N\geqslant 1$ and $N>p$ such a function can be written as
\beq \label{Taylor}
\pi(r) = \sum_{\ell=0}^{N-1} c_{2\ell} \, r^{2\ell} \, + \, \sum_{\ell=p}^{N-1} c_{2\ell+1} \, r^{2\ell+1} \, + \, q_{2N}(r) \, r^{2N},
\eeq
where $c_\ell = \pi^{(\ell)}(0)/\ell !$ and the remainder $q_{2N}(r)$ satisfies \\ $\lim_{r\to 0}q_{2N}(r) = 0$. 
In particular, for small $r$,
\beq  \label{Prop2}
\pi^\prime(r) - r \, \pi^{\prime\prime}(r) \, \sim \,  \left\{ 
\begin{array}{l l}
O(r^0)  ,  &  \text{if } \,  p =0,\\
O(r^2)  ,  &  \text{if } \,  p =1,\\
O(r^3)  ,  &  \text{if } \,  p \geqslant 2.
\end{array} \right .
\eeq

Let us now assume that the metric potentials are $p$-regular, with $p \geqslant 1$, and satisfy the conditions underlying~\eqref{Taylor}. It is straightforward to check that the components~\eqref{Riem} of the Riemann tensor have the following small-$r$ behaviour and regularity properties:
\beq \label{Hom0}
{]R_{\al\be\ga\de}[}  \, \sim \, r^0 , \quad \text{and} \quad \mathscr{P}({]R_{\al\be\ga\de}[}) = p - 1.
\eeq
Thus, the balanced components are $(p-1)$-regular. Since all the scalars $S^q_0 \in \mathscr{I}_{0}$ ($q\in\mathbb{N}$) have as building blocks the objects in~\eqref{Hom0}, it is clear that near the origin they tend to a constant value (they do not diverge) and that their first odd-order term occur at $r^{2p-1}$, for any $p \geqslant 1$, which means that they are $(p-1)$-regular---in short, $\mathscr{P}(S^q_0) = p-1 \, \forall q\in\mathbb{N}$. Hence, if $p=1$, the elements in $\mathscr{I}_{0}$ are 0-regular but not higher-order regular. Finally, it is easy to see that if the metric potentials are only 0-regular some components ${]R_{\al\be\ga\de}[}$ diverge, which means that the finiteness of the potentials is not enough to avoid curvature singularities. 

The next step is consider terms with one covariant derivative of the Riemann tensor. By direct calculation one can verify that the non-zero components satisfy
\beq \label{Hom1}
{]\na_\mu R_{\al\be\ga\de}[} \, \sim \, \left\{ 
\begin{array}{l l}
r^0  ,  &  \text{if } \,  p = 1 
\\
r^1  ,  &  \text{if } \,  p \geqslant 2. 
\end{array} \right .
\eeq
In the evaluation of some of these components it is necessary to use the identity~\eqref{Prop2}, which only contain terms at least of order $r^2$ for $p\geqslant 1$. This is an important feature, as the leading terms $r^0$ or $r$ would generate singular balanced components.

Thus, we can say that $r{]\na_\mu R_{\al\be\ga\de}[}$ is $(p-1)$-regular, which means that it is not $1$-regular for $p=1$. 
Of course, since the object in~\eqref{Hom1} has an odd number of indices, it is not possible to make any scalar with only one of it. However, considering an even number of them we can build scalars such as $(\na_\mu R_{\al\be\ga\de})^{2}\in \mathscr{I}_{2}$. 
From Eq.~\eqref{Hom1} we see that these scalars are regular for any $p \geqslant 0$, but they are only $(p-1)$-regular, just like the scalars in $\mathscr{I}_{0}$.

Having established the regularity order and the behaviour near $r=0$ for the quantities ${]R_{\al\be\ga\de}[}$ and $r{]\na_\mu R_{\al\be\ga\de}[}$ we are in position to extend considerations for a generic building block with any number of derivatives,
\beq \label{Component}
\na_r^{\ell} \na_\th^m \na_\phi^n R_{\al\be\ga\de}.
\eeq
Given two natural numbers $d$ and $k$, let us first assume the hypotheses:
\begin{itemize}
\item[$Ia$.] the total number $d=\ell+m+n$ of derivatives is \textit{odd}, and  $\mathscr{P}(r \, {]\na_r^{\ell} \na_\th^m \na_\phi^n R_{\al\be\ga\de}[})=k$ for some $k \geqslant 1$, for any combination of $\ell,m,n$ such that $\ell+m+n=d$;
\item[$Ib$.] there exists the limit $\lim_{r\to 0} {]\na_r^{\ell} \na_\th^m \na_\phi^n R_{\al\be\ga\de}[}$. Together with $Ia$, this is equivalent to $r \, {]\na_r^{\ell} \na_\th^m \na_\phi^n R_{\al\be\ga\de}[} \sim O(r^2)$.
\end{itemize}
We shall refer to the hypotheses above in the concise notation $I(d,k)$.
Now we take one more covariant derivative of~\eqref{Component}, analysing each case separately. Applying $\na_\phi$ we get:
\begin{equation} \label{DerPhi}
\begin{split}
& \na_\phi \na_r^\ell \na_\th^m \na_\phi^n R_{\al\be\ga\de} =  - \ell \Ga^\phi_{\phi r} \na_r^{\ell-1} \na_\th^m \na_\phi^{n+1} R_{\al\be\ga\de}  
\\ & \quad - n \Ga^r_{\phi\phi} \na_r^{\ell+1} \na_\th^m \na_\phi^{n-1} R_{\al\be\ga\de} - n \Ga^\th_{\phi\phi} \na_r^{\ell} \na_\th^{m+1} \na_\phi^{n-1} R_{\al\be\ga\de} 
\\
& \quad - m \Ga^\phi_{\phi\th} \na_r^{\ell} \na_\th^{m-1} \na_\phi^{n+1} R_{\al\be\ga\de} 
 - \Ga^\nu_{\phi\al} \na_r^\ell \na_\th^m \na_\phi^n R_{\nu\be\ga\de} 
 \\
 &\quad - \Ga^\nu_{\phi\be} \na_r^\ell \na_\th^m \na_\phi^n R_{\al\nu\ga\de} 
 - \Ga^\nu_{\phi\ga} \na_r^\ell \na_\th^m \na_\phi^n R_{\al\be\nu\de} 
 \\
 &\quad - \Ga^\nu_{\phi\de} \na_r^\ell \na_\th^m \na_\phi^n R_{\al\be\ga\nu}  .
\end{split}
\end{equation} 
Assuming $Ia$ and $Ib$ above and taking into account~\eqref{Chris} it follows, for example,
\beq
\lim_{r\to 0} {]\Ga^r_{\phi\phi} \na_r^{\ell+1} \na_\th^m \na_\phi^{n-1} R_{\al\be\ga\de}[} = \text{const.}
\nonumber
\eeq
and
\beq
\begin{split}
\mathscr{P}( {]\Ga^r_{\phi\phi} \na_r^{\ell+1} \na_\th^m \na_\phi^{n-1} R_{\al\be\al\be}[}) & = \mathscr{P}( r \, {] \na_r^{\ell+1} \na_\th^m \na_\phi^{n-1} R_{\al\be\ga\de}[}) - 1 
\nonumber
\\
& = k-1.
\end{split}
\eeq
All in all, it is straightforward to verify that analogous relations hold for every term in~\eqref{DerPhi}, giving
\beq \label{Resultado1ph}
{]\na_r^\ell \na_\th^m \na_\phi^{n+1} R_{\al\be\ga\de}[} \sim O(r^0) , 
%\quad \text{and} 
\quad \mathscr{P}( {]\na_r^\ell \na_\th^m \na_\phi^{n+1} R_{\al\be\ga\de}[}) = k - 1.
\eeq
Notice, however, that if in $Ia$ we had allowed $k=0$, then in $Ib$ we would have $r \, {]\na_r^{\ell} \na_\th^m \na_\phi^n R_{\al\be\ga\de}[} \sim O(r)$ and therefore ${]\na_r^\ell \na_\th^m \na_\phi^{n+1} R_{\al\be\ga\de}[}$ would not be regular.

Similar considerations can be applied to the terms \\ $\na_r^\ell \na_\th^{m+1} \na_\phi^n R_{\al\be\ga\de}$ and $\na_r^{\ell+1} \na_\th^m \na_\phi^n R_{\al\be\ga\de}$, with the same qualitative result of~\eqref{Resultado1ph}.  Thus, if $Ia$ and $Ib$ above hold, then 
\beq
\mathscr{P}({]\na_\mu \na_r^\ell \na_\th^m \na_\phi^n R_{\al\be\ga\de}[}) =  k-1
\eeq 
for any index $\mu$. 
One can use this relation, for example, to investigate the building blocks $\na_{\mu_1} \na_{\mu_2} R_{\al\be\ga\de}$, with two derivatives. We already know that if the metric potentials are $p$-regular, then $\mathscr{P}({] r \, \na_\mu R_{\al\be\ga\de}[}) = p - 1$. Therefore, it is immediate to get $\mathscr{P}({] \na_{\mu_1} \na_{\mu_2} R_{\al\be\ga\de}[}) = p - 2$, for $p \geqslant 2$. On the other hand,
if $p=1$ then  ${] \na_\mu R_{\al\be\ga\de}[} \sim O(r^0)$ (see~\eqref{Hom1}) and some components diverge, \textit{e.g.},  ${] \na_\phi \na_\phi R_{\al\be\al\be}[} \sim r^{-1}$. This explains why $\Box R$ diverges if $p=1$ (compare with the example in Sec.~\ref{Sec3}). If $p>1$, then ${] \na_{\mu_1} \na_{\mu_2} R_{\al\be\ga\de}[} \sim O(r^{0})$, but the occurrence of the first odd power depends on $p$; for $p=2$ it is already at linear order (thus this term is regular but not $1$-regular). To sum up, only if $p\geqslant 2$ then all the scalars in $\mathscr{I}_2$ are bounded at $r=0$.

After having established the result of applying one covariant derivative to a regular object with odd number of derivatives, let us now assume:
\begin{itemize}
\item[$II$.] the total number $d=\ell+m+n$ of derivatives is \textit{even}, and $\mathscr{P}( {]\na_r^{\ell} \na_\th^m \na_\phi^n R_{\al\be\ga\de}[})=k$ for some $k \geqslant 1$, for any combination of $\ell,m,n$ such that $\ell+m+n=d$,
\end{itemize}
which we shall refer as $II(d,k)$. 
Since the action of one covariant derivative converts an odd polynomial into even, it is more useful to consider the regularity order of the quantity $r\na_\mu \na_r^\ell \na_\th^m \na_\phi^n R_{\al\be\ga\de}$. As in the previous cases, it is necessary to take one covariant derivative with respect to each coordinate. Let us consider 
\beq \label{DerR2}
\begin{split}
& \na_r \na_r^\ell \na_\th^m \na_\phi^n R_{\al\be\ga\de}  =  \partial_r \na_r^\ell \na_\th^m \na_\phi^n R_{\al\be\ga\de}  
\\
& \quad - m \Ga^\th_{r\th} \na_r^{\ell} \na_\th^{m} \na_\phi^{n} R_{\al\be\ga\de}  - n \Ga^\phi_{r\phi} \na_r^{\ell} \na_\th^{m} \na_\phi^{n} R_{\al\be\ga\de}
\\
& \quad - \sum_{\varepsilon \in \mathcal{A}} \left[   \Ga^\th_{r\th}  \, \de(\varepsilon,\th) + \Ga^\phi_{r\phi} \, \de(\varepsilon,\phi) \right]  \na_r^{\ell} \na_\th^{m} \na_\phi^{n} R_{\al\be\ga\de}
\\
& \quad =  \left[  \partial_r  - \frac{m + n + j_\om}{r}\right]  \na_r^\ell \na_\th^m \na_\phi^n R_{\al\be\al\be},
\end{split}
\eeq
where $\mathcal{A} = \lbrace \al,\be,\ga,\de \rbrace$ and $j_\om$ is the number of angular indices ($\th$ and $\phi$) in $R_{\al\be\ga\de}$.
Since ${]\na_r^{\ell} \na_\th^m \na_\phi^n R_{\al\be\ga\de}[}$ is at least 1-regular, its limit as $r \to 0$ must be finite and it can be written as 
\beq
{]\na_r^{\ell} \na_\th^m \na_\phi^n R_{\al\be\ga\de}[} = c_0 + O(r^2) 
\eeq
for a constant $c_0$.
Also, recalling the definition of the balanced components,
\beq
\na_r^{\ell} \na_\th^m \na_\phi^n R_{\al\be\ga\de} = r^{m + n + j_\om} \left[ c_0 + O(r^2) \right] ,
\eeq
whence
\beq
\partial_r \na_r^{\ell} \na_\th^m \na_\phi^n R_{\al\be\ga\de} =  c_0 (m + n + j_\om) r^{m + n + j_\om - 1} 
+ O(r^{m + n + j_\om+1}) .
\nonumber
\eeq
Therefore, although
\beq
{]\partial_r \na_r^{\ell} \na_\th^m \na_\phi^n R_{\al\be\ga\de}[} = \frac{c_0(m + n + j_\om)}{r} + O(r)
\eeq
diverges, this singularity is precisely cancelled by the extra term appearing in~\eqref{DerR2}, so that
\beq \label{Resultado2r}
{r \,]\na_r^{\ell+1} \na_\th^m \na_\phi^n R_{\al\be\ga\de} [} \sim O(r^2) ,
\quad \mathscr{P}({r \, ]\na_r^{\ell+1} \na_\th^m \na_\phi^n R_{\al\be\ga\de}[})=k.
\eeq

The behaviour of~\eqref{Resultado2r} can be verified also for the terms involving an extra derivative $\na_\th$ or $\na_\phi$. In fact, the remaining components can be dealt by commuting the derivatives, applying the Bianchi identities and noticing that $\na_\th^{m^\prime} \na_\phi^{n^\prime} R_{\al\be\ga\de}=0$ if the total number of indices $\th$ (or $\phi$) is odd (for any $m^\prime,n^\prime$).
Therefore, if $II$ holds, then,
\beq \label{AbRes}
{r \,] \na_\mu \na_r^\ell \na_\th^m \na_\phi^n R_{\al\be\ga\de} [} \sim O(r^2) ,
\quad
\mathscr{P}(r{]\na_\mu \na_r^\ell \na_\th^m \na_\phi^n R_{\al\be\ga\de}[})=k.
\eeq
It is useful to notice that had we allowed $k=0$ in $II$, the only change in the result~\eqref{AbRes} is that ${r \,] \na_\mu \na_r^\ell \na_\th^m \na_\phi^n R_{\al\be\ga\de} [} \sim O(r)$.

Given the relations $II(d,k) \Longrightarrow I(d+1,k)$ and $I(d,k) \Longrightarrow II(d+1,k-1)$, starting from $R_{\al\be\ga\de}$ and $\na_\mu R_{\al\be\ga\de}$ one can successively apply covariant derivatives until one reaches $k=0$. The considerations above show that if the metric potentials $\chi_{0,2}$ are $(p+1)$-regular, 
all the terms ${]\na_r^\ell \na_\th^m \na_\phi^n R_{\al\be\ga\de}[}$ are bounded if $d = \ell + m + n \leqslant 2p+1$;
therefore, all the invariants in $\mathscr{I}_{2p}$ are regular, as the theorem stated.

%%%%%%%%%%%%%%%%%%%%%%%%%%%%%%

\end{document}